\documentclass[12pt, twocolumn]{aastex631}
\usepackage{graphicx} 
\usepackage{float}
\usepackage{booktabs}

\received{Aug 10, 2024}
\revised{Jan 9, 2025}
\accepted{Jan 29, 2025}

\begin{document}

\title{Characterising the multiple protostellar system VLA 1623$-$2417 with JWST, ALMA and VLA: outflow origins, dust growth and an unsettled disk}

\author[0009-0007-2837-8207]{Isaac C. Radley}
\affiliation{School of Physics and Astronomy University of Leeds, LS2 9JT, Leeds, UK}

\author[0000-0002-2189-6278]{Gemma Busquet}
\affiliation{Departament de Física Quàntica i Astrofísica (FQA), Universitat de Barcelona, Martí i Franquès 1, E-08028  Barcelona, Catalonia, Spain}
\affiliation{Institut de Ciències del Cosmos (ICCUB), Universitat de Barcelona, Martí i Franquès 1, E-08028 Barcelona,  Catalonia, Spain}
\affiliation{Institut d'Estudis Espacials de Catalunya (IEEC), Esteve Terradas 1, edifici RDIT, Parc Mediterrani de la Tecnologia (PMT) Campus del Baix Llobregat - UPC 08860 Castelldefels (Barcelona), Catalonia, Spain}

\author[0000-0003-1008-1142]{John D. Ilee}
\affiliation{School of Physics and Astronomy University of Leeds, LS2 9JT, Leeds, UK}

\author[0000-0003-2300-2626]{Hauyu Baobab Liu}
\affiliation{Department of Physics, National Sun Yat-Sen University, No. 70, Lien-Hai Road, Kaohsiung City 80424, Taiwan, R.O.C.}
\affiliation{Center of Astronomy and Gravitation, National Taiwan Normal University, Taipei 116, Taiwan}

\author[0000-0002-3972-1978]{Jaime E. Pineda}
\affiliation{Max-Planck-Institut f\"ur extraterrestrische Physik, Giessenbachstrasse 1, D-85748 Garching, Germany}

\author[0000-0001-7552-1562]{Klaus M. Pontoppidan}
\affiliation{Jet Propulsion Laboratory, California Institute of Technology, 4800 Oak Grove Drive, Pasadena, CA 91109, USA}

\author[0000-0003-1283-6262]{Enrique Mac\'ias}
\affiliation{European Southern Observatory, Karl-Schwarzschild-Str. 2, 85748, Garching bei München, Germany}

\author[0000-0002-7026-8163]{Mar\'ia Jos\'e Maureira}
\affiliation{Max-Planck-Institut für Extraterrestrische Physik (MPE), D-85748 Garching, Germany}

\author[0000-0001-9249-7082]{Eleonora Bianchi}
\affiliation{INAF, Osservatorio Astrofisico di Arcetri, Largo E. Fermi 5, 50125 Firenze, Italy}

\author[0000-0001-7491-0048]{Tyler L. Bourke}
\affiliation{SKA Observatory, Jodrell Bank, Lower Withington, Macclesfield, SK11 9FT, United Kingdom}

\author[0000-0003-1514-3074]{Claudio Codella}
\affiliation{INAF, Osservatorio Astrofisico di Arcetri, Largo E. Fermi 5, 50125 Firenze, Italy}

\author[0000-0001-8694-4966]{Jan Forbrich}
\affiliation{University of Hertfordshire, Centre for Astrophysics Research, College Lane, Hatfield, AL10 9AB, UK}

\author[0000-0002-3829-5591]{Josep M. Girart}
\affiliation{Institut d'Estudis Espacials de Catalunya (IEEC), Esteve Terradas 1, edifici RDIT, Parc Mediterrani de la Tecnologia (PMT) Campus del Baix Llobregat - UPC 08860 Castelldefels (Barcelona), Catalonia, Spain}
\affiliation{Institut de Ciències de l'Espai (ICE-CSIC), Campus UAB, Carrer de Can Magrans S/N, E-08193 Cerdanyola del Vallès, Catalonia}

\author[0000-0003-2684-399X]{Melvin G. Hoare}
\affiliation{School of Physics and Astronomy University of Leeds, LS2 9JT, Leeds, UK}

\author[0000-0002-4762-2240]{Ricardo Hernández Garnica}
\affiliation{Instituto de Radioastronomía y Astrofísica, Universidad Nacional Autonóma de México, Apartado Postal 3-72, Morelia 58090, Michoacán, Mexico}

\author[0000-0003-4493-8714]{Izaskun Jim\'enez-Serra}
\affiliation{Center of Astrobiology (CAB), CSIC-INTA, Ctra. de Ajalvir km 4, E-28850, Torrej\'on de Ardoz, Madrid, Spain}

\author[0000-0002-5635-3345]{Laurent Loinard}
\affiliation{Instituto de Radioastronomía y Astrofísica, Universidad Nacional Autonóma de México, Apartado Postal 3-72, Morelia 58090, Michoacán, Mexico}
\affiliation{Black Hole Initiative at Harvard University, 20 Garden Street, Cambridge, MA 02138, USA}
\affiliation{David Rockefeller Center for Latin American Studies, Harvard University, 1730 Cambridge Street, Cambridge, MA 02138, USA}

\author[0000-0001-7776-498X]{Jazmín Ordóñez-Toro}
\affiliation{Instituto de Radioastronomía y Astrofísica, Universidad Nacional Autonóma de México, Apartado Postal 3-72, Morelia 58090, Michoacán, Mexico}

\author[0000-0003-2733-5372]{Linda Podio}
\affiliation{INAF, Osservatorio Astrofisico di Arcetri, Largo E. Fermi 5, 50125 Firenze, Italy}

\begin{abstract}
Utilising JWST, ALMA and the VLA we present high angular resolution (0\farcs06--0\farcs42), multi-wavelength (4$\,\mu$m--3$\,$cm) observations of the VLA~1623-2417 protostellar system to characterise the origin, morphology and, properties of the continuum emission. JWST observations at 4.4\,\micron\ reveal outflow cavities for VLA~1623~A and, for the first time, VLA~1623~B, as well as scattered light from the upper layers of the VLA~1623~W disk. We model the millimetre-centimetre spectral energy distributions to quantify the relative contributions of dust and ionised gas emission, calculate dust masses, and use spectral index maps to determine where optical depth hinders this analysis. In general, all objects appear to be optically thick down to $\sim$90 GHz, show evidence for significant amounts (10's--100's $M_{\oplus}$) of large ($>1$\,mm) dust grains, and are dominated by ionised gas emission for frequencies $\lesssim\,$15\,GHz. In addition, we find evidence of unsettled millimetre dust in the inclined disk of VLA~1623~B possibly attributed to instabilities within the circumstellar disk, adding to the growing catalogue of unsettled Class 0/I disks. Our results represent some of the highest resolution observations possible with current instrumentation, particularly in the case of the VLA. However, our interpretation is still limited at low frequencies ($\lesssim22$\,GHz) and thus motivates the need for next-generation interferometers operating at centimetre wavelengths.
\end{abstract}

\section{Introduction}

The primary mode of planet formation occurs via the successive coagulation and subsequent growth of small ($\sim$0.1\,\micron) ISM-like dust particles through 12 orders of magnitude to planetary scales \citep[e.g.][]{ Weidenschilling1977}.  Traditionally this was thought to be a slow process, and therefore confined to the later stages of evolution in protoplanetary disks, in particular the Class II phase (with ages $\sim10^{5.5} - 10^7$ years, \citealt{Pascucci22}).  However, recent theoretical advances have shown that the planet formation process can proceed much more quickly than originally anticipated, for example by processes such as the streaming instability \citep[see][for a review]{Johansen2014}.  This, coupled with recent observational evidence of possible planet-induced substructure in young disks \citep[e.g.][]{Segura-Cox2020}, has led to a renewed interest in studying planet formation in the earliest stages of star formation, for example with recent surveys such as FAUST \citep[Fifty AU STudy of the chemistry in the disc/envelope system of solar-like protostars,][]{Codella21}, eDISK \citep[Early Planet Formation in Embedded Disks,][]{eDISK} and CAMPOS \citep[ALMA Legacy survey of Class 0/I disks in Corona australis, Aquila, chaMaeleon, oPhiuchus north, Ophiuchus, and Serpens,][]{Hsieh24}.

As a result of the dust growth processes previously mentioned, we expect to find a distribution of dust grain sizes in protoplanetary disks.  Since thermal emission arises at wavelengths related to the dust grain size, i.e.\ $a_{max} \sim \frac{\lambda}{2\pi}$, this means we must utilise observational wavelengths across a similarly broad range in order to fully characterise the dust population in disks.  Scattered light has been used to trace $\sim$$\mu$m size dust grains, which are coupled to the gas and suspended in the atmospheres of disks \citep[e.g.][]{Villenave19,Ginski24} providing constraints on their vertical extent.  Longer wavelength observations in the (sub)-mm regime can begin to trace larger dust that becomes decoupled from the gas, settling toward the planet-forming midplane of the disk \cite[e.g.][]{Ansdell16,Andrews_2018}.  Until recently it was assumed such observations gave an unobstructed view of the raw material available for planet formation.  However, even at these longer wavelengths, significant optical depths can still be reached in the disks due to the high density of material \citep{Zhu2019ApJ...877L..18Z,Chung2024ApJS..273...29C}.  Moving toward the longest wavelengths in the cm regime provides the best chance at mitigating any effects of high optical depth \citep[see, e.g.][]{CarrascoGonzalez19}.  Unfortunately, this is complicated by the fact that these observations do not only trace thermal dust emission, but are also sensitive to emission from ionised gas through winds, jets and magnetospheric processes operating in young stellar objects (YSOs) and their circumstellar disks \citep[see e.g.][]{Liu2014ApJ...780..155L,Coutens19}.  

This contribution of ionised gas emission can have significant effects on the derived spectral energy distribution (SED) at low frequencies \citep[e.g. $\lesssim$30\,GHz,][]{Dzib13} with a shape that is unique to the properties driving the emission. We can consider two general regimes for which ionised gas mechanism is primarily contributing based on the derived SED and therefore spectral index, $\alpha$, where F$_\nu \propto \nu^{\alpha}$. In the first case, emission arises from free-free interactions in an ionised gas which can be located close to the central star such as at the base of a jet, or as a photoevaporative disk wind. In this regime we expect $\alpha$ values between $-0.1$ and +1 \citep{Pascucci12,Macias16,Anglada18} with a typical value of $\sim$\,0.6 for an idealised jet \citep[see e.g.][]{Reynolds86}. Alternatively, we may also have significant contributions from magnetospheric activity such as gyrosynchrotron emission \citep{Feigelson85} where free electrons interact with magnetic field lines originating from the central star giving rise to a non-thermal emission process with a typical $\alpha \sim -0.7$ for stars within the ISM \citep{Condon16}. It is important to note that within star forming regions, the higher density environment can lead to variations in the observed spectral index value with previous surveys finding $\alpha < -0.1$ for possible non-thermal emitters \citep[e.g.][]{Dzib13} and theoretical predictions demonstrating values potentially as low as --2 \citep[e.g.][]{Dulk1985}. While these values in theory offer a potential discriminant between emission mechanism, the actual observed $\alpha$ will vary due to its dependence on optical depth and local environment.  Therefore, we must obtain high spatial resolution and high sensitivity observations of YSOs across multiple characteristic wavelengths in order to be able to properly decompose the relative contributions of each of these processes to the observed emission.

\begin{figure*}
    \centering
    \includegraphics[width=\textwidth]{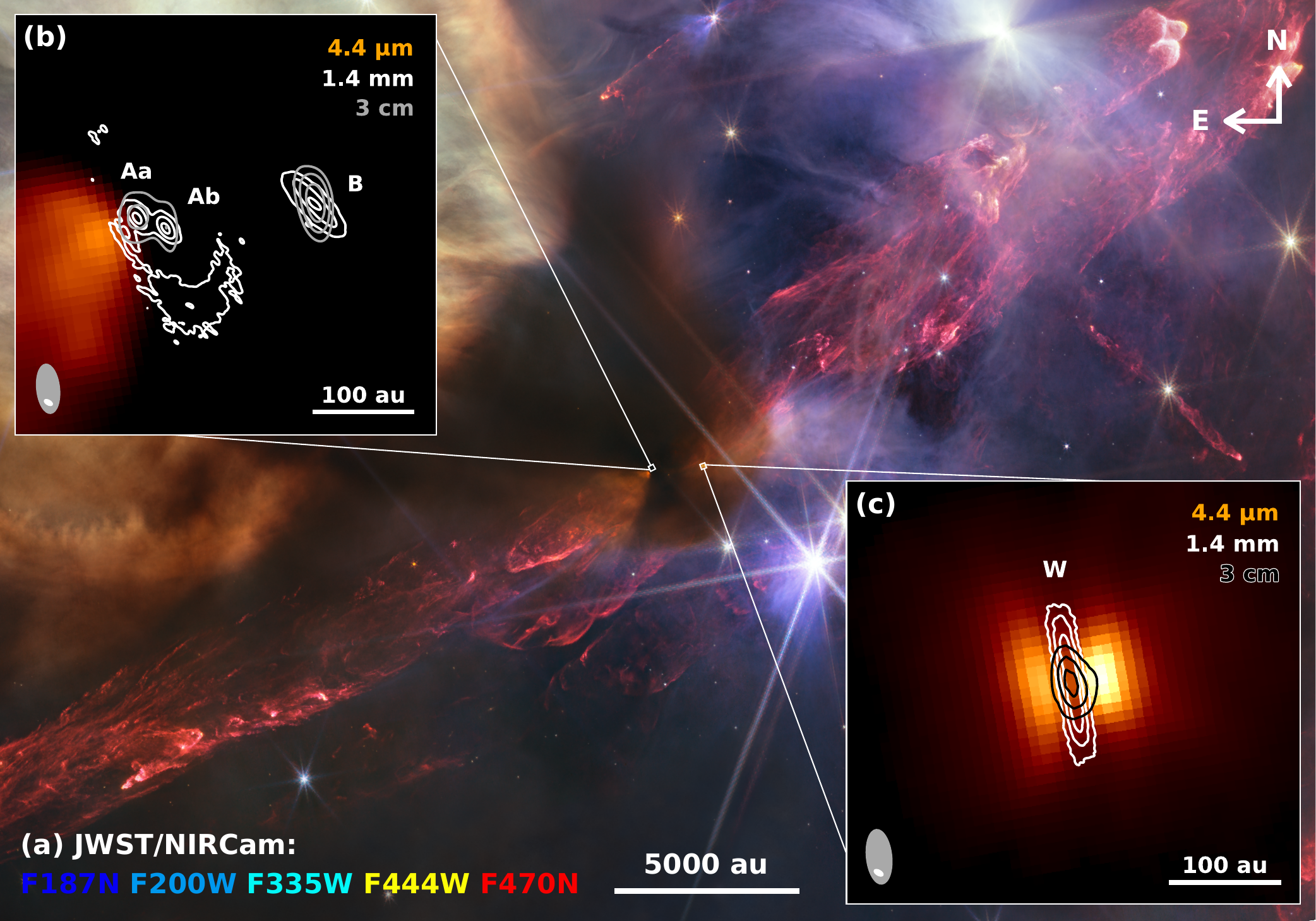}
    \caption{{\bf(a)} JWST/NIRCam First Anniversary image centred on the VLA 1623 system, showing the pc-scale bipolar outflows (NASA/ESA/CSA/STScI).  {\bf (b)} Zoom-in toward VLA 1623\,Aa, Ab and B.  The colour scale shows emission from the JWST/NIRCam 4.4\,\micron\ filter (peak normalised with a square root stretch) tracing the base of the SE outflow.  White contours show the ALMA 1.4\,mm (217\,GHz) emission dominated by thermal dust emission (8, 125, 350$\sigma$, where $\sigma=0.02$\,mJy\,beam$^{-1}$).  Grey contours show the VLA 3\,cm (10\,GHz) emission (7, 12, 20$\sigma$, where $\sigma=0.01$\,mJy\,beam$^{-1}$) tracing a combination of thermal dust and ionised gas emission.  {\bf (c)} Zoom-in toward VLA\,1623\,W.  Colour scale as in (a), tracing scattered light from the upper layers of the edge-on disk.  White contours show 1.4\,mm emission (8, 25, 45$\sigma$) and black show 3.0\,cm emission (3, 5, 7$\sigma$), which similarly trace thermal dust and a combination of thermal dust and ionised gas, respectively.}
    \label{fig:overview}
\end{figure*}

The VLA~1623-2417 system (hereafter VLA~1623) is located in the Ophiuchus~A (L1688) star forming region. Ophiuchus is one of the closest star forming regions at 138.4 pc \citep{OrtizLeon2018} making it an excellent observational candidate for both maximising spatial resolution as well as sensitivity to ensure faint multiwavelength emission can be detected above the noise level. Due to these favourable characteristics, both the individual YSOs and properties of the region as a whole have been extensively studied \citep[e.g.][]{Ridge2006,Dzib13, Friesen2017, ODISEA1}.

The VLA~1623 system is now known to be comprised of four objects -- Aa, Ab, B, and West (W)\footnote{Aa and Ab have also been referred to as A1 and A2 respectively, see e.g. \citet{Ohashi22}.}. The proto-binary system, VLA~1623~A (Aa + Ab), is considered to be the prototypical Class 0 object \citep{Andre93} and has been found to be surrounded by a circumbinary disk \cite[see e.g.][]{Murillo13_CBDisk,Hsieh20,Sadavoy24}. Previous observations \citep[see e.g.][]{Murillo13} classified VLA~1623~A as a single object due to their limited angular resolution until \citet{Harris18} and \citet{Kawabe18} were able to successfully resolve the object into two protostellar components separated by 0.21\arcsec \citep[$\sim30$\,au,][]{Harris18}. In contrast, the nature of VLA~1623~B and VLA~1623~W has previously been subject to debate.  Both objects were thought to possibly be shocked cloudlets \citep{Bontemps97} instead of YSOs. However, recent observations from (sub-)mm dust continuum \citep[][]{Harris18,Michel22} and kinematic gas tracers \citep[][]{Ohashi22,Mercimek23,Codella24} point towards embedded, disk-bearing protostars. VLA~1623~B has been suggested to be a Class 0 YSO \citep{Murillo18} observed to have an inclined protostellar disk \citep[$i>70\degr$,][]{Sadavoy2019,Sadavoy24} and is separated by $\sim1\farcs2$ ($\sim170$\,au) from its companion VLA~1623~A \citep{Harris18}.  VLA~1623~W is found at a much larger separation to A \citep[$\sim\,11\arcsec$ $\sim\,1500$\,au,][]{Mercimek23} and is thought to be a slightly more evolved Class I YSO with a nearly edge-on protostellar disk \citep[$i\sim75-80\degr$,][]{Sadavoy2019,Michel22,Sadavoy24}.  

Previous studies of the VLA 1623 system found a large-scale, bipolar outflow in CO and H$_2$ originating from VLA~1623~A \citep[see e.g.][]{Andre1990,Davis1995,Dent1995,Caratti2006}. More recently the VLA~1623~A outflow has also been detected in multiple molecular tracers and forbidden lines e.g. c-C$_3$H$_2$ \citep{Murillo_18_chem} and [\ion{O}{1}] \citep{Nisini15}. Spitzer \citep{Spitzer} observations at 4\,\micron\ revealed a cavity aligned with the outflow in the proximity of VLA~1623~A \citep[see e.g.][]{Zhang2009,Kawabe18}. Interestingly, VLA~1623~B is theorised to have jet driving a collimated outflow \citep[see e.g.][]{Santangelo2015} however no outflow cavity has been detected close to the protostar. Due to their deeply embedded nature \citep[see e.g.][]{Looney2000, Murillo13}, neither VLA~1623~A nor B protostars are directly detected in these Spitzer images.  Conversely, VLA~1623~W appears as a point-like infrared source which can been clearly seen in the \citet{Kawabe18} 4.5\,\micron\ image as well as at 8\,\micron\ and 24\,\micron\ in \citet{Murillo13}.

Unfortunately, due to the proximity of VLA~1623~A and B, the origin of detected outflows has been difficult to attribute to one or multiple objects. Observations of CO(3--2) from \citet{Hsieh20} point towards two outflows being driven by A and B individually whereas observations of $^{12}$CO(2--1) in \citet{Hara21} indicate that both outflows may originate from the Aa and Ab in the VLA~1623~A protobinary. Therefore, there is a degree of ambiguity on which specific objects are driving the outflows detected in the VLA 1623 system.

In this paper we present high resolution Very Large Array (VLA), Atacama Large Millimetre Array (ALMA) and James Webb Space Telescope (JWST) continuum observations of the VLA~1623 system. We describe our observational set-up and final imaging parameters in Section \ref{sec:Observations}. In Section \ref{sec:Results} we present our continuum images, fluxes and derived spectral indices for each object. Additionally, through SED analysis, we probe possible interpretations of the dust and ionised gas populations. We discuss our findings in Section \ref{sec:discussion} and finally Section \ref{sec:conclusion} gives an overview of our conclusions as well as contextualising this work in regards to an upcoming wider survey of Ophiuchus expanding on the work of \citet[][]{Coutens19}.

\section{Observations and Data Reduction}\label{sec:Observations}

Figure~\ref{fig:overview} shows an overview of the L1688 region observed with JWST, overlaid with our VLA and ALMA observations of the VLA 1623 system, demonstrating the connection of these YSOs to the wider star forming environment.  Below we discuss the observational set up and data reduction steps taken for each of the instruments.

\subsection{JWST/NIRCam}

A section of the Ophiuchus star-forming region was observed by the near infrared camera (NIRCam) instrument \citep{Rieke23} on the 7$^{\text{th}}$ March, 5$^{\text{th}}$ April and 6$^{\text{th}}$ April 2023 to produce the first anniversary image for the James Webb Space Telescope (JWST, PID: 2739, P.I. Pontoppidan). The field covers $\sim$49 square arcminutes and includes the VLA 1623 outflow, as well as several well-studied young stars and protostars. Six filters were used: F115W, F187N, F200W, F335M, F444W, and F470N, which trace the \ion{H}{1} Paschen $\alpha$ line, the 3.3\,$\mu$m PAH band, scattered light and H$_2$S(9) rotational line and CO fundamental rovibrational band, respectively.
The observation was designed and processed generally following the procedures outlined in \cite{Pontoppidan2022}. Briefly, a 3$\times$2 mosaic was constructed with 71.5\% overlap in rows to produce an image with uniform depth. Each tile was observed using the FULLBOX+6TIGHT dither pattern resulting in a maximum exposure time of 1416-1674\,s for the broad and medium-band filters and 2834\,s for the narrow-band filters. The images were processed using the JWST calibration pipeline \citep{Bushouse23} version 11.16.21 with CRDS context \texttt{jwst$\_$1077.pmap}. The image resolution at 2.0 $\mu$m corresponds to 0.06'' (or 7 au at the distance of Ophiuchus).

\subsection{ALMA}

We obtained archival ALMA observations of the VLA 1623 region (project code 2019.1.01074.S\footnote{We note here to avoid confusion that the ALMA Science Archive incorrectly records the source name as VLA 16293 despite this field covering the position of VLA 1623.}, P.I. Maureira) across both Band 6 (217 GHz, \citealt{Ediss2004}) and Band 3 (93 GHz, \citealt{Claude2008}).  The Band 6 observations were taken on 7$^{\text{th}}$ August, 3$^{\text{rd}}$ October, and 21$^{\text{st}}$ October 2021 in configuration C43-8 for a total of 87 minutes with a phase centre of 16$^h$26$^m$26$\rlap{.}{}^{s}42$ -24\degr24\arcmin30\farcs00 (J2000). The observations had a precipitable water vapour (PWV) between 0.8--1.5\,mm and between 41 and 45 antennas depending on execution.  Baseline lengths ranged from 70\,m to 11.6\,km corresponding to maximum recoverable scale of 0.85\arcsec. 

The correlator was set to frequency division mode (FDM) with spectral windows across 216--234\,GHz.  Pipeline calibration was carried out by the UK ALMA Regional Centre (ARC) node, supplemented by additional self-calibration with CASA (version 6.4.1, \citealt{CASA}). A continuum visibility measurement set was created by combining all channels from the continuum spectral windows with a central frequency of 217\,GHz (1.4\,mm) and total bandwith of 5.6\,GHz after flagging any channels that may contain strong line emission (although none was apparent).  Three rounds of phase only self-calibration were performed with solution intervals of infinity, 30s and 18s, respectively.  Model images were created using masks that encompassed strong emission.  This resulted in an increase of peak signal-to-noise (SNR) of a factor of $\sim$3.  A final round of amplitude self-calibration was attempted but did not result in an improvement of SNR or image fidelity.  

The Band 3 observations were taken on 07 September 2021 in configuration C43-9/10 for a total of 85 minutes with a PWV between 3.1-3.3\,mm and a phase centre of 16$^h$26$^m$26$\rlap{.}{}^{s}$42 -24\degr24\arcmin30\farcs00 (J2000).  Baseline lengths ranged from 122\,m to 16.2\,km corresponding to a maximum recoverable scale of 0.86\arcsec\ with 49 antennas. The correlator was set to FDM mode with spectral windows across 92--107\,GHz.  Pipeline calibration was carried out by the UK ALMA ARC node.  The central frequency of the final continuum visibility measurement was 93\,GHz (3.2\,mm) with a bandwidth of 7.5\,GHz.  As with the Band 6 observations, pipeline calibration was followed by three rounds of phase only self calibration (with solution intervals of infinity, 30s, and 18s) providing an increase in peak SNR of factor $\sim$4.

For both Band 6 and 3, imaging was performed with \texttt{tclean} using Briggs weighting and multi-scale, multi-frequency synthesis (with scales of 0, 5, 15, 30 times the synthesised beam). Our fiducial image products, used in the subsequent analyses, have beam sizes of 0\farcs06$\times$0\farcs03 [60\degr] and 0\farcs07$\times$0\farcs05 [57\degr] for Band 6 and 3, respectively (outlined fully in Table \ref{tab:obs_params}). Finally, all image products were corrected for the primary beam response using CASA \textit{impbcor}.

\subsection{VLA}

We carried out observations of the L1688 region with the Karl G.\ Jansky Very Large Array (VLA) of the National Radio Astronomy Observatory, using the Q- (44\,GHz), K- (22\,GHz) and X- (10\,GHz) bands (project code 22A-164, P.I. Busquet) with phase centres \mbox{16$^h$26$^m$27$\rlap{.}{}^{s}$80} \mbox{-24\degr24\arcmin40\farcs30}, \mbox{16$^h$26$^m$25$\rlap{.}{}^{s}$63} \mbox{-24\degr24\arcmin29\farcs4} and, \mbox{16$^h$26$^m$24$\rlap{.}{}^{s}$60} \mbox{-24\degr23\arcmin37\farcs00} (J2000), respectively.  The observations were taken across two different epochs (2$^{\text{nd}}$ and 7$^{\text{th}}$ April 2022) with the array in the A configuration and a maximum baseline of $\sim$36~km. Our maximum baseline correlates to a maximum recoverable scale of 1\farcs2 , 2\farcs4 and 5\farcs3 for the Q-, K- and X-band observations respectively. The data was taken using the 3-bit samplers and with two 2~GHz (X-band) and four 2~GHz wide basebands (K- and Q-bands) and in full polarization. 

The data was processed using the VLA Calibration Pipeline\footnote{https://science.nrao.edu/facilities/vla/data-processing/pipeline} within CASA (version 6.5.2). We additionally applied one round of phase-only self-calibration to the Q- and K-band observations, 2 rounds for the X-band data collected on the 2$^{\text{nd}}$ April and 4 rounds for the X-band data from the 7$^{\text{th}}$ April. Our model images were created using masks which encompassed strong emission. In both the Q- and K-band observations we combined spectral windows and scans with an infinite solution interval for the former and down to 90 s for the latter. For both X-band observations we started with an infinite solution interval (the entire scan) and then down to 96 s for April 2$^{\text{nd}}$ while for April 7$^{\text{th}}$ we went down to 48s and then 24s. Self-calibration marginally improved the peak SNR by a factor of $\sim$1.2 and 1.5 for the Q- and X-band epochs however SNR remained similar for the K-band observations. For the purposes of this work, self-calibration mainly acts to significantly improve the image fidelity for all sources across the respective fields of view by reducing contaminating sidelobes.

As we are combining multiple epoch observations (2$^{\text{nd}}$ and 7$^{\text{th}}$ April 2022) we also must assess the level of variability of each source (discussed in detail in section \ref{sec:Variability}). In order to do this, we create images using CASA \textit{tclean} \citep{CASA} from both epochs using the hogbom deconvolver \citep[see e.g.][]{Hogbom} and a Briggs robust weighting of 2 to maximise sensitivity. The beamsizes of the variability images used in this analysis are 0\farcs11$\times$0\farcs1 [28\degr] and 0\farcs11$\times$0\farcs08 [8\degr] for Q-band;  0\farcs26$\times$0\farcs14 [26\degr] and 0\farcs2$\times$0\farcs18 [17\degr] for K-band and, 0\farcs43$\times$0\farcs19 [7\degr] and 0\farcs42$\times$0\farcs22 [6\degr] for X-band on the 2$^{\text{nd}}$ and 7$^{\text{th}}$ April, respectively. We present these variability image parameters in Table \ref{tab:obs_params}. 

For the rest of the analysis within this paper we make use of fiducial images derived from combining the measurement sets of the 2$^{\text{nd}}$ and 7$^{\text{th}}$ April 2022 observations. For our fiducial VLA~1623~A and B image products we use a Briggs robust weighting of 0.5 for all VLA frequencies. The fiducial beam sizes for VLA~1623~A and B are 0\farcs09$\times$0\farcs06 [17\degr], 0\farcs18$\times$0\farcs12 [19\degr], 0\farcs35$\times$0\farcs16 [6\degr] for the Q-, K- and X-bands, respectively. For VLA~1623~W we use separate images of the same field using a robust weighting of 2 for the K and X-band images and a $uv$-taper of $0\farcs2$ for the Q-band image. The fiducial beam sizes for VLA~1623~W are 0\farcs26$\times$0\farcs14 [8\degr], 0\farcs22$\times$0\farcs16 [24\degr], 0\farcs42$\times$0\farcs21 [7\degr] for the Q-, K- and X-bands, respectively. VLA~1623~W is also positioned close to the edge of the primary beam ($\sim16\arcsec$) in the 44\,GHz observations and so naturally has an increased noise level (0.07\,mJy\,beam$^{-1}$) compared to VLA~1623~AB (0.04\,mJy\,beam$^{-1}$, see Table \ref{tab:master}), however this does not hinder our analysis. Our fiducial imaging parameters, presented in Table \ref{tab:obs_params}, were chosen to provide an optimal trade off between image fidelity, SNR and angular resolution. Finally, we correct all images for the primary beam response using CASA \textit{impbcor}, ensuring accurate fluxes can be measured.

\begin{figure*}
    \centering
\includegraphics[width=\textwidth]{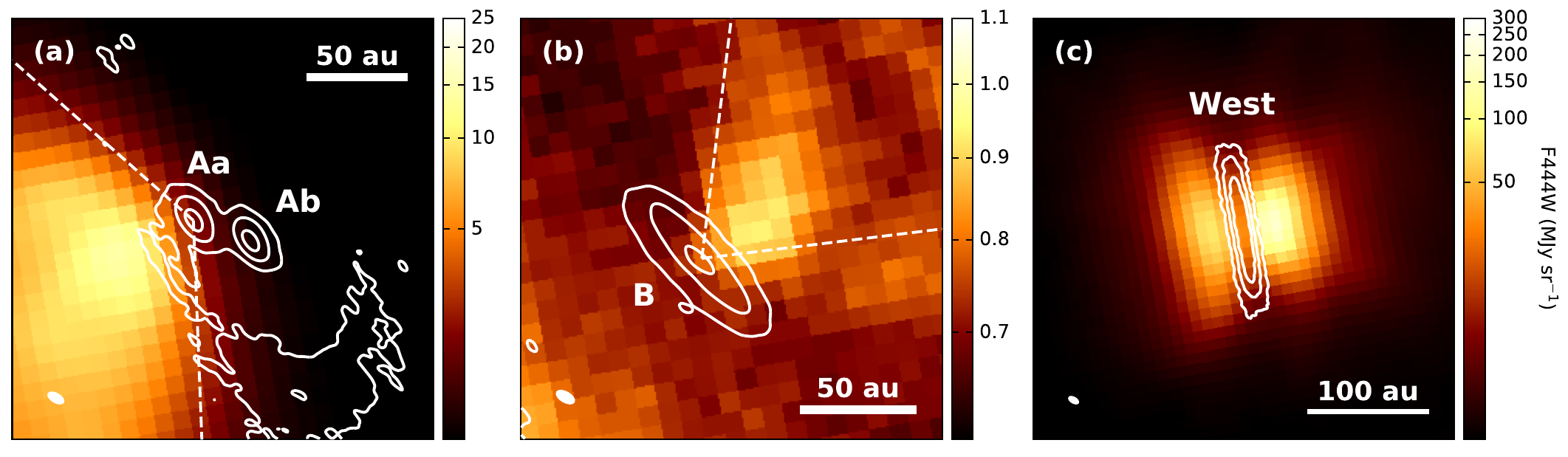}
    \caption{JWST/NIRCam F444W (4.4\,\micron) images toward VLA~1623~Aa/Ab, B, and W overlaid with ALMA 217\,GHz continuum emission (white contours at 8, 125, 350$\sigma$, where $\sigma=0.02$\,mJy\,beam$^{-1}$).  Each panel is peak normalised and shown with a logarithmic stretch.  Opening angles of 135\degr and 75\degr are shown for reference in panels (a) and (b), respectively.}
    \label{fig:F444W}
\end{figure*}

\subsubsection{Assessing variability}\label{sec:Variability}

Emission at low frequencies ($\lesssim\,$50\,GHz) from YSOs can be the result of a variety of physical processes which has been shown to result in variability on short (weeks to months) timescales \citep[see, e.g.,][]{Pech10,Dzib13,Liu2014ApJ...780..155L,Ubach2017,Coutens19}.  Before combining the VLA observations from the two different epochs (2$^{\text{nd}}$ and 7$^{\text{th}}$ April 2022), we investigated whether there was any significant variation in total flux.  We initially imaged the 10, 22 and 44\,GHz VLA measurement sets from each date with a Briggs robust value of 2 in order to maximise sensitivity.  We then used the Python Blob Detector and Source Finder package \citep[PyBDSF,][]{PyBDSF}, a Gaussian decomposition algorithm which takes into account the local noise field and any point spread function (PSF) effects to accurately determine both the morphology and total flux of sources in interferometric images.  Our measured fluxes, extracted from a single Gaussian fit, are reported in Table \ref{tab:EpochFluxes}.  We then applied the variability criterion from \citet{DiazMarquez24} which uses both the rms and calibration uncertainties of the minimum and maximum observed flux as an indication of variability between each epoch, and compares this to the maximum flux deviation.  Using this, we found that VLA observations of VLA~1623~Aa, Ab, B and W can be considered non-variable to the 3$\sigma$ level, and so we are able to combine the measurement sets from each epoch for further imaging.

\subsection{Astrometric alignment}
\label{sec:astrometric}

Our use of three different observational instruments necessitated an investigation into the relative astrometric alignment before we could confidently compare their image products.  We found that the JWST/NIRCam and ALMA astrometry agreed extremely well, matching object positions on a sub-pixel scale (e.g.\ less than 30\,mas).  For all data presented here, we rotate the JWST images to the orientation of the ALMA images using the \texttt{reproject} package\footnote{\url{https://reproject.readthedocs.io}}.  

When comparing the ALMA and VLA data we encountered non-constant offsets between the positions of objects across the fields, likely induced by the self-calibration process on the VLA data.  To correct this, we consider a small box region ($\sim 0.1\arcsec\times0.1\arcsec$) around each object for each consecutive frequency pair considered (e.g. 93\,GHz and 44\,GHz). We then smooth the image with the smallest beamsize to the largest beamsize of the frequency pair ensuring a common angular resolution between both images. We fit a 2D gaussian to both images using PyBDSF and perform an initial shift in the image plane using \textit{scipy.ndimage.shift} to adjust the lower frequency data to the position reported in the higher frequency image. We then refine this shift using an image difference minimisation processes in which we explore a grid of sub-pixel shifts around this value, comparing the absolute value of the relative difference between the images. Our final shifts are chosen based on those which minimise the relative differences between the two images (presented in Appendix \ref{sec:APXAstrometry}). We further check the implemented shift by visually inspecting the overlap between the 3-, 5- and 10-$\sigma$ contours of each image to ensure they are co-spatial.

\section{Results \& Analysis}\label{sec:Results}

In the following sections we discuss the initial results and analysis performed on each of the datasets, and how the observations of each object at different wavelengths compare with one another.    

\subsection{Morphology \& Fluxes}

\subsubsection{JWST/NIRCam}\label{sec:JWST_images}

Due to the highly embedded nature of the VLA~1623 system \citep[A$_{\rm V}\sim40$ mag,][]{Wilking1989} previous infrared detections of the VLA~1623~AaAb and B protostars were unfeasible. However, due to its separation from AB, VLA~1623~W has been previously detected and appears as a bright infrared source at 4.5\,\micron\ and 8\,\micron\ \citep[see e.g.][]{Murillo13,Kawabe18}. We therefore concentrate on the JWST/NIRCam F444W (4.4\,\micron) images, which are shown in Figure~\ref{fig:F444W}.

The brightest emission is detected toward VLA~1623~W, where we see two prominent lobes on either side of a darker lane (traced by the 217\,GHz emission) toward the position of the protostellar disk.  There is a brightness asymmetry between the two lobes, with the dimmer toward the East suggesting this represents the far side of the disk that is being partially obscured with respect to the near side. Alternatively, due to the morphology of this asymmetry, the Eastern edge may actually be the nearside of the disk viewed through a more extincted line of sight. These properties are very similar to observations of scattered light toward other edge-on disks observed in both infrared and millimetre images, for example Tau 042021 \citep{Duchene2024}.  This scattered light emission toward VLA~1623~W originates from high in the disk atmosphere, up to approximately 50~au from the midplane.     

Much fainter emission is detected toward both VLA~1623~Aa and B (approximately factors of ten lower in each case) with an asymmetric conical morphology appearing to originate from close to the central regions of each protostar. The emission morphology seen from VLA~1623~Aa in panel (a) has been previously detected in Spitzer 4\,\micron\ images \citep[see e.g.][]{Zhang2009,Murillo13,Kawabe18} and aligns with outflows detected in molecular emission lines such as CS(5-4), C$_2$H \citep{Ohashi22} and c-C$_3$H$_2$ \citep{Murillo_18_chem} as well as forbidden lines such as [\ion{O}{1}] \citep{Nisini15}. This emission is therefore attributed to an outflow cavity. The asymmetry is likely caused by a projection effect as the conical morphology aligns with previously observed, blue-shifted, $^{12}$CO(2-1) emission \citep[see e.g.][]{Hara21}. Therefore, we are likely seeing the near side of the outflow which is less extincted than the far side.

In regards to panel (b), this is the first detection of an outflow cavity originating from VLA~1623~B. \citet{Santangelo2015} had previously compared their high velocity CO(2-1) observations with [\ion{O}{1}] observations from \citet{Nisini15} and theorised that B may be launching its own jet. We find that the 4.4\,\micron\ outflow cavity is aligned with the expected direction of this collimated jet. Furthermore, they theorised that this jet component will be much more compact than the jet associated with VLA~1623~A.

Dashed lines overlaid on panels (a) and (b) in Figure~\ref{fig:F444W} show opening angles of 135\degr\ and 75\degr\ for VLA~1623~A and B respectively, fitted by-eye to encompass the observed emission. We report the angle interior to these dashed lines as our estimate for the opening angle of each object. Interestingly, the smaller opening angle of VLA~1623~B corroborates the highly collimated jet hypothesis of \citet{Santangelo2015} mentioned above. For both VLA~1623~Aa and B, the 4.44\,\micron\ emission appears to be tracing the base of outflowing material along directions perpendicular to the circumstellar disks.  Our measured opening angle for VLA~1623~Aa of 135\degr is significantly wider than outflow opening angles measured from observations of sub-mm molecular line emission (e.g. 20--30\degr, \citealt{Dent1995,Hara21}). This may be due to the very high sensitivity of the JWST/NIRCam observations, or due to the fact that observations of molecular species only trace specific excitation conditions.

\subsubsection{ALMA and VLA}\label{sec:FluxAndGallery}
\begin{figure*}[htbp]
    \centering
    \includegraphics[width=0.85\textwidth]{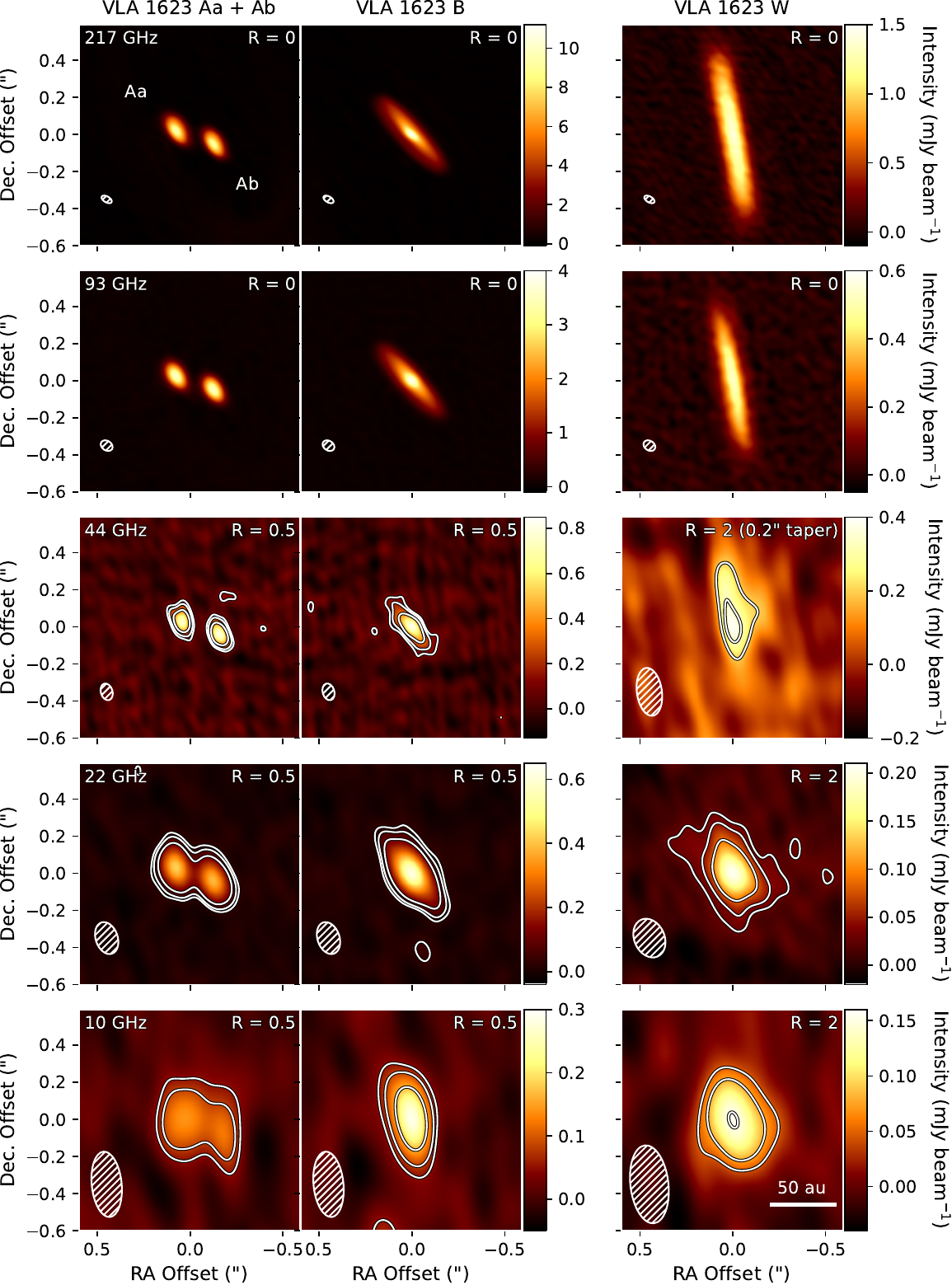}
    \caption{Continuum images of VLA 1623 Aa + Ab (Left), B (Middle) and W (Right). We show 3-, 5-, and 10-$\sigma$ contours based on the image RMS in white solid lines for all VLA images. The $\sigma$ values used in each VLA image for each object are shown as the $\sigma_{\rm rms}$ presented in Table \ref{tab:master}. The beam size is shown as the white hatched ellipse in the bottom left of each image. The Briggs robust of each row is shown in the top right of each panel and the frequency of each row is shown in the top left of each panel in the first column. We additionally indicate a 50 au scale bar in the bottom right panel, shared across all images.}
    \label{fig:Gallery}
\end{figure*}

Figure~\ref{fig:Gallery} shows the self-calibrated images of VLA 1623 A, B and W. As in Section \ref{sec:Variability}, we utilise PyBDSF to measure integrated fluxes, (deconvolved) sizes and position angles by fitting 2D gaussians to each object from each image. We also estimate the local noise level, $\sigma_{\rm rms}$, using CASA \textit{imstat} by measuring the RMS around each object. Finally, we estimate the inclination of each object by considering the ratio of major ($\theta_{\text{maj}}$) and minor ($\theta_{\text{min}}$) axis sizes at each frequency i.e. $i=\text{arccos}\left(\frac{\theta_{\text{min}}}{\theta_{\text{maj}}}\right)$. This inclination calculation assumes the disk to be geometrically thin and therefore may underestimate the true inclination. We present all discussed values in Table \ref{tab:master}.  

\begin{deluxetable*}{lcccccccc}[ht]
\tablecaption{Fluxes, size and orientation of each object measured from the ALMA and VLA observations.
\label{tab:master}}
\tablehead{
\colhead{Object} &
\colhead{Frequency} &
\colhead{F$_{\rm int}$} &
\colhead{$\sigma_{\rm rms}$}&
\colhead{F$_{\rm Dust}$} &  
\colhead{f$_{\rm Ionised}$} &
\colhead{Deconvolved Size} &
\colhead{PA} &
\colhead{Inclination}
\\
\colhead{} &
\colhead{GHz} &
\colhead{mJy} & 
\colhead{mJy beam$^{-1}$}&
\colhead{mJy} &
\colhead{$\%$} &
\colhead{au} &
\colhead{$\degr$} &
\colhead{$\degr$}
}
\startdata
VLA~1623~Aa & 217 & 45 $\pm$ 5 & 0.05 & 45 & 0.8 & 14.73 $\times$ 8.21 [ 0.08 $\times$ 0.04 ] & 30.9   & 56 \\
& 93 & 9.0 $\pm$ 0.5 & 0.01 & 8.7 & 3.2 & 13.02 $\times$ 6.58 [ 0.08 $\times$ 0.04 ] & 30.8   & 60 \\
& 44 & 1.2 $\pm$ 0.2 & 0.04 & 1.0 & 17.6 & 11 $\times$ 7.1 [ 1 $\times$ 0.5 ] & 15.7   & 50 \\
& 22 & 0.46 $\pm$ 0.05 & 0.01 & 0.29 & 37.3 & 8.9 $\times$ 4.3 [ 0.8 $\times$ 0.3 ] & 172.5   & 61 \\
& 10 & 0.23 $\pm$ 0.05 & 0.02 & 0.10 & 55.9 & 29 $\times$ 8 [ 7 $\times$ 4 ] & 110.8   & 73 \\
\hline
VLA~1623~Ab & 217 & 40  $\pm$  4 & 0.05 & 39.6 & 0.1 & 13.61 $\times$ 6.96 [ 0.08 $\times$ 0.03 ] & 28.2   & 59 \\
 & 93 & 8.8  $\pm$  0.4 & 0.01 & 8.7 & 0.5 & 12.12 $\times$ 6.26 [ 0.07 $\times$ 0.04 ] & 30.1   & 59 \\
 & 44 & 1.2 $\pm$ 0.2 & 0.04 & 1.2 & 4.3 & 10.8 $\times$ 5.4 [ 0.9 $\times$ 0.4 ] & 31.5   & 60 \\
 & 22 & 0.44  $\pm$  0.05 & 0.01 & 0.38 & 14.1 & 10.9 $\times$ 4.9 [ 0.9 $\times$ 0.4 ] & 36.2  & 63 \\
 & 10 & 0.12 $\pm$  0.03 & 0.02 & 0.04 & 63.1 & --- & ---  & --- \\
\hline
VLA~1623~B & 217 & 107  $\pm$  11 & 0.05 & 107 & 0.2 & 38.8 $\times$ 10.89 [ 0.1 $\times$ 0.03 ] & 41.9  & 74 \\
 & 93 & 22 $\pm$ 1 & 0.01 & 21 & 0.8 & 36.7 $\times$ 9.08 [ 0.1 $\times$ 0.03 ] & 41.6  & 76 \\
 & 44 & 2.4 $\pm$  0.3 & 0.04 & 2.3 & 6.8 & 23 $\times$ 2.1 [ 1 $\times$ 0.3 ] & 42.5  & 85 \\
 & 22 & 1.0 $\pm$ 0.1 & 0.01 & 0.9 & 16.0 & 25.6 $\times$ 1.8 [ 0.8 $\times$ 0.2 ] & 41.5 & 86 \\
 & 10 & 0.35  $\pm$  0.04 & 0.02 & 0.19 & 45.4 & 26 $\times$ 8 [ 4 $\times$ 1 ] & 30.8 & 71 \\
 \hline
VLA~1623~W & 217 & 52  $\pm$  5 & 0.03 & 51.7 & 0.1 & 80.5 $\times$ 12.2 [ 0.8 $\times$ 0.1 ] & 10.0  & 81 \\
 & 93 & 8.2  $\pm$  0.4 & 0.01 & 8.1 & 0.9 & 83 $\times$ 11.0 [ 1 $\times$ 0.2 ] & 10.2  & 82 \\
 & 44 & 0.9  $\pm$  0.2 & 0.07 & 0.8 & 11.0 & 77 $\times$ 5 [ 19 $\times$ 2 ] & 8.5  & 87 \\
 & 22 & 0.37  $\pm$  0.05 & 0.01 & 0.2 & 33.9 & 35 $\times$ 18 [ 3 $\times$ 1 ] & 28.8  & 59 \\
 & 10 & 0.24 $\pm$  0.04 & 0.02 & 0.08 & 67.5 & 35 $\times$ 14 [ 7 $\times$ 4 ] & 75.9 & 66 \\
\bottomrule
\enddata
\tablecomments{Integrated fluxes ($F_{\rm int}$) and local noise level, $\sigma_{\rm rms}$, are derived as discussed in Section \ref{sec:FluxAndGallery}. Dust fluxes ($F_{\rm dust}$) and ionised gas fraction ($f_{\rm ion}$) are derived in Section \ref{sec:SED}. All integrated fluxes have been corrected for the primary beam response. Deconvolved disk diameters are derived from the FWHM of our PyBDSF fits with uncertainties in square brackets. We derive inclinations using the ratio of major to minor disk sizes as described in Section \ref{sec:FluxAndGallery}}

\end{deluxetable*}

Due to the high resolution of our images (0\farcs06--0\farcs4, see Table \ref{tab:obs_params}) we manage to successfully resolve VLA 1623 A into its binary components Aa and Ab at all frequencies, although they are only marginally resolved at 10\,GHz. By calculating the average difference in peak pixel position of Aa and Ab we find they are separated by $0.219 \pm 0.003$ arcseconds, or $\sim$ 30\,au. Since we resolve Aa and Ab at ALMA frequencies and (marginally) at 44\,GHz, these are our best insights into the morphological nature of these objects. Both objects share a similar deconvolved position angle of $30.9 \pm 0.5$\degr\ and $28.2 \pm 0.5$\degr\ for Aa and Ab respectively, with a much lesser dust continuum radial extent than their companion B (see Table \ref{tab:master}). At 217\,GHz we also partially detect the AaAb circumbinary disk at the 5$\sigma$ level (see Figure~\ref{fig:F444W}a), although the scale of the disk \citep[$\sim2\farcs2$,][]{Harris18} is beyond our largest angular scale for this array configuration. This implies that the emission will be filtered, and so we do not analyse this aspect further. The two binary components are morphologically similar across all frequencies except at 10 GHz where Aa appears to dominate Ab in size/extent. Both objects (Aa, Ab) have similar fluxes across our observations varying on average by around 5\% from each other excluding 10\,GHz where we see the largest flux difference between the two objects with Aa having almost double the flux of Ab. 

We see that the morphology of VLA~1623~B shows indications of being a highly inclined disk ($i=74\degr$, Table \ref{tab:master}), most notably at 217\,GHz where a dark lane appears to obscure emission from the central source (see Section \ref{sec:B_EmSurf} for further discussion). Previous works \citep[see e.g.][]{Harris18,Sadavoy24} have found similarly high inclination angles for VLA~1623~B. However, at 44 and 22~GHz this disk structure is much more dominated by a compact 10$\sigma$ central component with a fainter (3$\sigma$) emission surface enclosing the core. VLA~1623~B lies approximately 160\,au from the midpoint of Aa and Ab.

VLA~1623~W is located $\sim$1300\,au from VLA~1623~B. This source shows the most significant morphological change across wavelengths with the higher frequencies (217\,GHz and 93\,GHz) showing an elongation in the North-South direction and the lower frequencies (22 GHz and 10 GHz) dominated by a compact central core. Unfortunately, in our 44\,GHz observation, the image fidelity of VLA~1623~W is partially hindered by increased noise due to the object's position at the edge of the primary beam. However, we can still clearly see a compact central component, similar to the other VLA frequencies. We detect a strong compact component at 5$\sigma$ and a weaker 3$\sigma$ elongation towards the North similar to the 217 and 93 GHz images, possibly indicating the beginning of a transitional regime dominated by the central star. Additionally, there is a further (tentative, $\sim 2.5-3 \sigma$) elongation to the West which is shared by all three VLA observations but not detected at higher frequencies.

\subsection{Constructing millimetre-centimetre spectral energy distributions}

In order to examine the nature of the emission from each object we construct SEDs across the millimetre and centimetre wavelength regimes. To supplement our ALMA and VLA observations we incorporate ALMA Band 7 (350 GHz) data from \citet{Harris18}. We take the fluxes reported in \citet{Harris18} derived from images with angular resolutions similar to our own (i.e. 0.14\arcsec$\times$0.12\arcsec\ for AB and 0.2\arcsec$\times$0.17\arcsec\ for W) to ensure we are tracing emission on similar spatial scales. Further to this, we include additional 6\,GHz (C-band) observations from \citet{Dzib13}. The wide bandwidth of these observations allowed \citet{Dzib13} to measure fluxes from the upper and lower sub-bands, providing individual measurements at 7.5 and 4.5\,GHz, respectively. It is important to note that VLA~1623~Aa and Ab are not resolved into their binary components at either C sub-band frequency. We therefore consider the fluxes from the upper and lower C sub-bands as an upper limit for both binary components since we do not have any information regarding what share of this flux originates from Aa or Ab. The constructed SEDs for each object are shown in Figure~\ref{fig:SED} and Figure~\ref{fig:SingPopSED}. We summarise the fluxes used in each SED in Table \ref{tab:Fluxes}.

\subsubsection{Quantifying the dust and ionised gas spectral indices}\label{sec:SED}

\begin{figure*}
    \centering
    \includegraphics[width=0.7\textwidth]{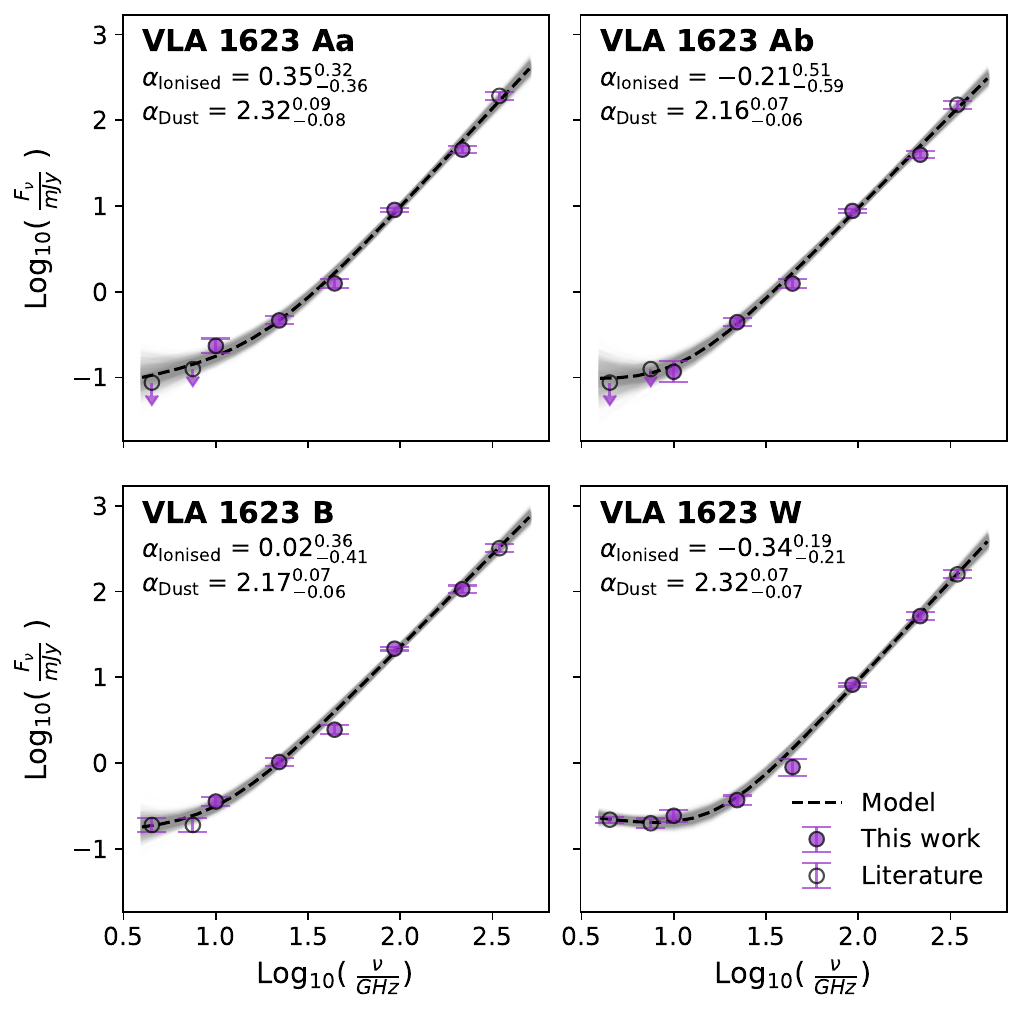} 
    \caption{SED of each object overlaid with the models described in Section \ref{sec:SED}.  Fluxes derived from this work are shown as filled circles while literature values are shown with open circles. Upper limits are shown as arrows.  The shaded region shows random draws from the posterior probability distribution given by the fitting procedure, and the black dashed line indicates the best fit.  Each panel also shows the high ($\alpha_{\text{Dust}}$) and low ($\alpha_{\text{Ionised}}$) frequency spectral indices derived from this fit along with their 16$^{th}$ and 84$^{th}$ percentile errors.}
    \label{fig:SED}
\end{figure*}

Using all fluxes from Table \ref{tab:Fluxes}, we implement a Markov Chain Monte Carlo (MCMC) approach using \textit{emcee} \citep{Emcee} to estimate the spectral indices arising from contributions of ionised gas ($\alpha_{\text{Ionised}}$) and dust ($\alpha_{\text{Dust}}$) to the observed flux from our SED. We make use of a double power law model of the form:

\begin{equation}\label{eq:DoublePowerLaw}
    F_\nu = A_1\left(\frac{\nu}{10\text{\,GHz}}\right)^{\alpha_{\text{Ionised}}}+A_2\left(\frac{\nu}{10\text{\,GHz}}\right)^{\alpha_{\text{Dust}}},
\end{equation}
where $F_\nu$ is the flux at frequency $\nu$ and $A_1$,\,$A_2$ are the scaling amplitudes. Using this two component model allows us to decompose the relative contributions of both dust thermal emission (dominant at the highest frequencies in the Rayleigh-Jeans regime) and ionised gas emission (dominant at the lowest frequencies).  We use a Gaussian likelihood and flat priors, $\vec{\theta}$, with the following constraints:
\begin{equation}
    P(\, \vec{\theta} \, ) =\left\{
    \begin{array}{ll}
    1, & \text{if } 0 < A_1 < 1.4. \\
    1, & \text{if } -1.5 < \alpha_{\text{Ionised}} <  1.5. \\
    1, & \text{if } 0 < A_2 < 1.4.  \\
    1, & \text{if } 1 < \alpha_{\text{Dust}} < 3.8.  \\
    -\infty, & \text{Otherwise.}
    \end{array}
    \right.
\end{equation}

Our amplitudes ($A_1$\,,\,$A_2$) were loosely constrained through a preliminary least squares fit using \textit{scipy.curvefit} and by definition must be non-zero. We expect $\alpha_{\text{Dust}}$ to vary between $\sim$ 3.8 if no dust growth is present \citep[see e.g.][]{Draine06} and 1.5 in the case of significant dust self-scattering \citep[see e.g.][]{Sierra20}. Therefore, we constrain $\alpha_{\text{Dust}}$ to lie within the 3.8--1.5 range, ensuring we are probing primarily dust thermal emission. We use the lower limit of $\alpha_{\text{Dust}}$ as the upper limit of $\alpha_{\text{Ionised}}$ as we want to ensure our value is not tracing the dust thermal emission. We expect ionised gas emission to have typical spectral indices between 0.7 and $-0.1$ for jets or winds and $-0.7$ for magnetospheric gyrosynchrotron emission \citep{Pascucci12,Condon16,Anglada18} and so we use a generous lower limit of $-1.5$ for $\alpha_{\text{Ionised}}$ to ensure we explore the maximum parameter space.

We use 32 walkers with 2$\times10^5$ steps to sample our distribution and remove a burn-in of 1500 allowing us to determine the $\alpha_{\text{Ionised}}$ and $\alpha_{\text{Dust}}$ spectral indices as well as their associated uncertainties for each object as shown in Figure~\ref{fig:SED}.

For all objects, we find $\alpha_{\text{Dust}}$ values between 2.1 and $\sim\,$2.3. VLA~1623~Ab and B have the lowest dust thermal spectral indices of the system with $\alpha_{\text{Dust}}$ values of 2.16 and 2.17 respectively. Every object has a very well constrained $\alpha_{\text{Dust}}$ with the highest associated uncertainty being roughly $\pm\,0.1$. As mentioned previously, $\alpha_{\text{Dust}}$ values between 3.8 and 2 indicate that some degree of dust growth may be present in these disks. The fact that all objects have $\alpha_{\text{Dust}}$ values close to 2 also points towards these disks being potentially optically thick.

Considering $\alpha_{\text{Ionised}}$, we find values between $-0.35$ and 0.4 for all objects. VLA~1623~W has the most negative ionised gas spectral index ($\alpha_{\text{Ionised}} = -0.34$) as well as the lowest uncertainty compared to the other objects. If we consider the uncertainties of the ionised gas spectral index, the minimum $\alpha_{\text{Ionised}}$ of VLA~1623~W ($\sim-0.6$) lies close to the expected spectral index for magnetospheric gyrosynchrotron emission \citep[$\sim-0.7$,][]{Condon16}. However, $\alpha_{\text{Ionised}}$ is also consistent with $-0.15$ within errors and thus we may be detecting a combination of optically thin and optically thick ionised gas emission. Additionally, we see a flattening of $\alpha_{\text{Ionised}}$ for VLA 1623 B, however, this only spans the upper (7.5\,GHz) and lower (4.5\,GHz) sub-bands of the C-band observation and so may hint at a change of regime at longer wavelengths. 

We note that in these MCMC runs we have not considered the 4.5\,GHz and 7.5\,GHz fluxes of Aa and Ab as strict upper limits as the implementation of such constraints within this kind of fitting procedure is non-trivial.  Therefore, some of the draws from the posterior distributions violate the upper limits, and the associated negative uncertainties in  $\alpha_{\text{Ionised}}$ for these objects are overestimated.  Nevertheless, we see that Aa has the highest $\alpha_{\text{Ionised}}$ (0.35) of all objects yet with a significant error of $\pm\,0.3$ this could also indicate either a flat spectral index consistent with zero or a slightly positive spectral index of 0.6--0.7. As discussed in the introduction, a spectral index between $-0.1$ and 1 may evidence jet or photoevaporative wind emission, however an idealised jet is expected to exhibit a spectral index of $\sim 0.6$ \citep{Reynolds86}. If we consider that the 4.5\,GHz and 7.5\,GHz observations are an upper limit then logically the spectral index must be either positive or flat unless Aa emits significantly at 4.5\,GHz. 

Considering VLA~1623~Ab, we find that although $\alpha_{\text{Ionised}} = -0.21^{+0.51}_{-0.59}$, the uncertainties are so high that the object could equally fall into all three spectral index regimes - positive, negative or flat. Once again, using our knowledge that the 4.5\,GHz and 7.5\,GHz observations are an upper limit, as well as considering that Ab has approximately half the flux of Aa in the X-band, it is not unlikely that it is similarly weaker at frequencies $\lesssim$\,10\,GHz and therefore the true spectral index may be higher than what we report here.

Finally, we can correct our measured fluxes for contaminating emission by using the ionised gas and dust thermal components of our SED fit to estimate the uncontaminated flux from thermal dust emission. To determine this, we evaluate Equation \ref{eq:DoublePowerLaw} assuming the dust thermal component has no contribution (i.e. $A_{2} = 0$) to estimate our ionised gas flux and subtract this from our integrated fluxes. We call this contaminant-corrected flux the dust flux, $F_{\rm Dust}$ and include it in Table \ref{tab:master} along with an estimate of the fraction of ionised emission, $f_{\rm Ionised}$, compared to the total integrated flux, $F_{\rm int}$.

\subsubsection{Reproducing the observed SEDs with combinations of dust and ionised gas} 
\label{sec:SED_dustmodel}

In addition to the dust and ionised gas spectral indices of these objects we can attempt to model the underlying population of dust and nature of the ionised gas mechanisms which create the resultant SEDs. Unfortunately, we do not have sufficient spatial resolution for frequencies $\lesssim 22$\,GHz to accurately describe the morphology of the dust disk for each object across our entire data range. In addition, in Class 0/I objects, a number of case studies have shown that in the high-density regions traced by VLA observations, the thermal structures may be complicated due to viscous heating (\citealt{Liu2019ApJ...884...97L,Liu2021ApJ...914...25L,Zamponi2021MNRAS.508.2583Z,Xu2022ApJ...934..156X,Xu2023ApJ...954..190X,Takakuwa2024ApJ...964...24T}).  Therefore, instead of the commonly adopted approach of fitting a passive disk model (e.g., \citealt{Chiang1997ApJ...490..368C}), we follow the approach of \citet{Liu2019ApJ...884...97L} and \citet{Liu2021ApJ...923..270L} to model the spatially-unresolved SED with components of dust and ionised gas emission.  For simplicity, we assume that the physical properties within each of these emission components are uniform. 

Since the dust emission in our objects may be optically thick at these frequencies (discussed in Section ~\ref{sec:SED}), the formulation needs to take into account the mutual obscuration of the emission components. For each source, the integrated flux density is expressed by 
\begin{equation}\label{eqn:multicomponent}
    F_{\nu} = \sum\limits_{i} F_{\nu}^{i} e^{-\sum\limits_{j}\tau^{i,j}_{\nu}}, 
\end{equation}
where $F_{\nu}^{i}$ is the flux density of the dust or ionised gas emission component $i$, and $\tau^{i,j}_{\nu}$ is the optical depth of the emission component $j$ to obscure the emission component $i$.

When considering the contribution of ionised gas emission we can use the analytical expression of $F_{\nu}^{i}$ and $\tau^{i,j}_{\nu}$ which depends on three free parameters: the electron temperature ($T_{\text{e}}$), the emission measure (EM), and the solid angle ($\Omega_{\text{ff}}$) and is provided in Appendix~\ref{appendix:freefree}.
Since we do not possess sufficient independent measurements at low frequencies to fully constrain these three parameters, we nominally assume $T_{\text{e}}=8000$\,K for all ionised gas emission components in our models.

To accurately model the dust emission within circumstellar disks, we must consider the effect of dust scattering (\citealt{Kataoka2015ApJ...809...78K,Liu2019ApJ...877L..22L,Zhu2019ApJ...877L..18Z}) and so to incorporate this into our model we use the DSHARP dust opacity and Equations (10)--(20) presented in \citet{Birnstiel2018ApJ...869L..45B} when modelling the dust emission components, which has previously been applied to Class 0 and Class I YSOs \citep[see e.g.][]{Takakuwa2024ApJ...964...24T,Guerra-Alvarado24}.  Specifically, for dust components that have temperatures below 170\,K we assume that they are coated with water ice and use the DSHARP-default opacity table which assumed that the dust grains are composed of water ice \citep{Warren1984ApOpt..23.1206W}, astronomical silicates \citep{Drain2003ARA&A..41..241D}, troilite, and refractory organics (\citealt{Henning1996A&A...311..291H}; see Table 1 of \citealt{Birnstiel2018ApJ...869L..45B}).  For dust components with temperatures greater than 170\,K  we use the corresponding DSHARP opacity for ice-free grains. The adopted sublimation temperature of 170\,K is motivated by the dependence of volatile sublimation on local gas pressure \citep[][their Equations 3 and 4]{Okuzumi2016ApJ...821...82O}. Previous studies have typically used sublimation temperatures in the range of 150-170\,K \citep[e.g. 160\,K,][]{Tobin2023}. However, due to the comparatively higher densities associated with Class 0/I objects, we choose the higher end of the temperature range. Further to this, it is assumed the dust grains are morphologically compact (i.e. low porosity),  consistent with recent ALMA observations (e.g. \citealt{Tazaki2019ApJ...885...52T}).  

The size-averaged dust opacities were evaluated by assuming a power law grain size ($a$) distribution $n(a) \propto a^s$ with $s=-3.5$  between the minimum and maximum grain sizes ($a_{\text{min}}$, $a_{\text{max}}$) (i.e., $n(a)=0$ elsewhere), which are insensitive to the value of $a_{\text{min}}$.  We artificially set $a_{\text{min}}=0.1$ $\mu$m, which is common in the studies of protoplanetary disks (e.g. \citealt{Chung2024ApJS..273...29C}).  The remaining free parameters for each dust emission component are dust temperature ($T_{\text{dust}}$), dust column density ($\Sigma_{\text{dust}}$), solid angle ($\Omega_{\text{dust}}$), and $a_{\text{max}}$.

\begin{figure*}[h]
    \centering
    \includegraphics[width=\linewidth]{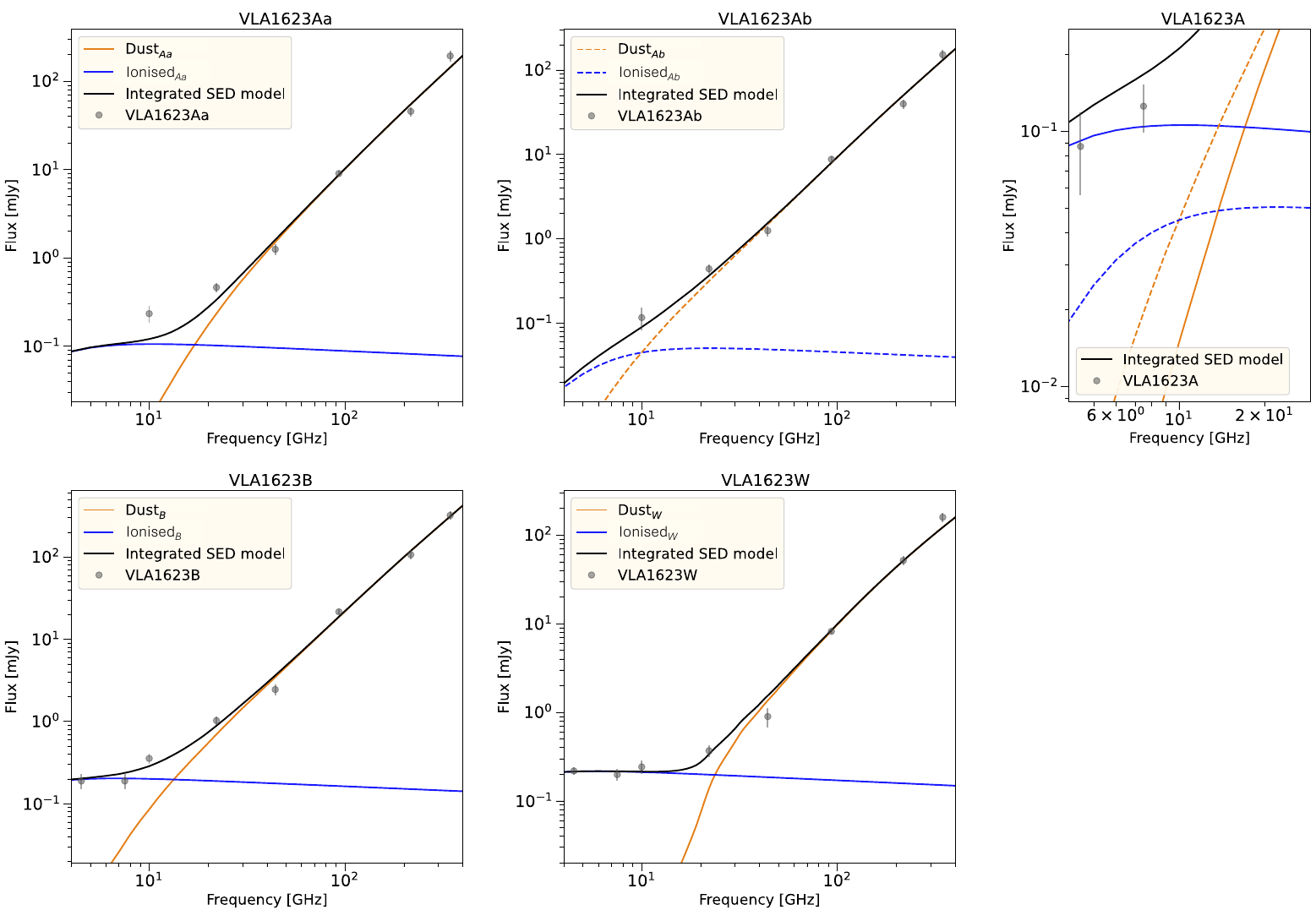}
    \caption{\textit{Top:} SED models for the VLA~1623~Aa and Ab resolved, integrated fluxes alongside the combined model, A, for the unresolved C-band data \citep{Dzib13}. Symbols show the observational data, coloured lines show the SEDs of individual emission components in our model (Table \ref{table:modelsingledust}). For clarity we show the Aa emission components as solid lines and Ab components as dashed lines in the combined model. Integrated flux densities are shown in black lines. \textit{Bottom:} Same as the top but for resolved observations of VLA~1623~B and W respectively. 
    }
    \label{fig:SingPopSED}
\end{figure*}

Using the assumptions and parameters above, we model both dust and ionised gas contributions to the SED of each object with an MCMC approach, again, using \textit{emcee}. We incorporate flat priors, allowing each parameter to increase or decrease by 1 order of magnitude. When considering VLA~1623~Aa and Ab, we fit the resolved 217-10\,GHz SED individually. However, for the 7.5\,GHz and 4.5\,GHz observations we use the integrated SED from both Aa and Ab models as these objects are unresolved at these frequencies. Our MCMC fitting used 300 walkers with 5$\times10^4$ steps and removed a burn in of 1$\times10^4$ allowing us to fully probe the parameter space. We present our best fit parameters and their respective uncertainties in Table~\ref{table:modelsingledust} as well as the resulting model SEDs in comparison to the observations in Figure~\ref{fig:SingPopSED}.

\begin{deluxetable*}{ lccccccr }
\tablecaption{Model parameters 
\label{table:modelsingledust}}
\tablewidth{800pt}
\tablehead{
\colhead{Component} &
\colhead{$T$} &
\colhead{$\Omega$\tablenotemark{a}} &
\colhead{$\Sigma_{\mbox{\tiny dust}}$} &  
\colhead{$a_{\mbox{\tiny max}}$} &
\colhead{EM} &
\colhead{$M_{\mbox{\tiny dust}}$}
\\
\colhead{} &
\colhead{(K)} &
\colhead{($10^{-12}$ sr)} & 
\colhead{(g\,cm$^{-2}$)} &
\colhead{(mm)} &
\colhead{(cm$^{-6}$\,pc)} &
\colhead{$M_{\oplus}$}
\\
 & (1) & (2) & (3) & (4) & (5) &  (6)
} 
\startdata
\multicolumn{7}{c}{\textbf{VLA~1623~Aa}} \\
Dust  & 218$^{+90}_{-43}$ & 0.24$^{+0.066}_{-0.072}$ & 4.5$^{+1.8}_{-1.3}$ & 17$^{+10}_{-6.0}$ & $\cdots$ &   28$^{+22}_{-14}$ \\
\\
Ionised gas  & 8000 & 4.3$^{+3.4}_{-1.6}\cdot$10$^{-2}$ & $\cdots$ & $\cdots$ & 2.98$^{+2.15}_{-1.53}\cdot10^{7}$ &  $\cdots$ \\
\\
\multicolumn{7}{c}{\textbf{VLA~1623~Ab}} \\
Dust  & 236$^{+95}_{-63}$ & 0.20$^{+0.08}_{-0.06}$ & 20$^{+14}_{-8.4}$ & 19$^{+15}_{-9.9}$ & $\cdots$ &   113$^{+82}_{-45}$ \\
\\
Ionised gas  & 8000 & 4.6$^{+6.2}_{-2.5}\cdot$10$^{-3}$ & $\cdots$ & $\cdots$ & 1.5$^{+1.8}_{-1.1}\cdot10^{8}$ &  $\cdots$ \\
\\
\multicolumn{7}{c}{\textbf{VLA~1623~B}} \\
Dust  & 113$^{+48}_{-29}$ & 1.1$^{+0.40}_{-0.34}$ & 23$^{+15}_{-9.1}$ & 15$^{+14}_{-8.9}$ & $\cdots$ &650$^{+810}_{-380}$ \\
\\
Ionised gas  & 8000 & 0.20$^{+0.14}_{-0.079}$ & $\cdots$ & $\cdots$ & 1.2$^{+0.97}_{-0.54}\cdot10^{7}$ &  $\cdots$ \\
\\
\multicolumn{7}{c}{\textbf{VLA~1623~W}} \\
Dust & 25$^{+5.5}_{-4.5}$ & 2.6$^{+0.79}_{-0.60}$ & 6.7$^{+13}_{-5.0}$ & 2.8$^{+7.2}_{-1.4}$ & $\cdots$  & 448$^{+1300}_{-360}$ \\
\\
Ionised gas  & 8000 & 0.28$^{+0.14}_{-0.082}$ & $\cdots$ & $\cdots$ & 8.8$^{+4.6}_{-3.2}\cdot10^{6}$ &  $\cdots$ \\
\\
\enddata
\tablecomments{
\tablenotemark{a} 1 sr $\sim$\,4.25$\times$10$^{10}$ square arcsecond.
(1) Dust temperature for dust emission components and electron temperature for ionised gas emission component. (2) Solid angle of the emission components. (3) Dust column density. (4) Maximum dust grain size. (5) Emission measure for the ionised gas emission component. (6) Dust mass in units of Earth masses ($M_{\oplus}$).
}
\end{deluxetable*}

Using a single dust population, we find that all objects are consistent with a maximum grain size $>1$\,mm. This result corroborates the findings of Section~\ref{sec:SED} where $\alpha_{\text{Dust}}$ has values between 3.8 and 2 for all objects which can be explained by the presence of grain growth in the circumstellar disk \citep[see e.g.][]{Draine06}. While these values are rather unconstrained, they represent the necessity of large dust grains in order to replicate the observed SED at low frequencies. 
This is particularly pertinent due to our inclusion of ionised gas in the model which contributes significantly at low frequencies \citep{Dzib13}. Therefore, we require not only contributions from the ionised gas, but also from large, hot dust grains.

Nevertheless, we find several issues arise from employing the single dust population and ionised gas model. For example, the $\Sigma_{\rm dust}$ and $a_{\rm max}$ have to have high values in order to reproduce the observed low spectral indices at $\sim$10--50 GHz. The very high $a_{\rm max}$ values here are not consistent with the previous detection of the 872 $\mu$m polarization due to dust self-scattering in these sources (\citealt{Harris18}). Furthermore, in all four sources, the single dust component model cannot reproduce both the low spectral indices at $\sim$10--44 GHz and the high spectral indices at 217--350 GHz simultaneously.

For a better explanation of the observational data it is plausible to consider the next level of complexity by including more emission components and the mutual obscuration of these components. A natural way to replicate the more optically thick spectral indices ($\sim$2) for frequencies $\gtrsim$22 GHz without contributing too much flux at $\sim$6 GHz, is to include another dust emission component that has a high $a_{\rm max}$ value and a high temperature, which is obscured at $\gtrsim$30 GHz. This may indicate dust mass segregation or dust growth, which have been previously observed in Class 0/I protoplanetary disks \citep[see e.g.][]{Han2023}.

For all four sources, the discrepancy between model and data at $\gtrsim$\,200 GHz can be amended by including a relatively optically thin dust emission component. Unfortunately, the $a_{\rm max}$ of this relatively optically thin dust emission component cannot be constrained by the present data. However, previous detections of dust self-scattering at 872 $\mu$m indicate that the $a_{\rm max}$ of these relatively optically thin dust components are of the order of $\sim$100 $\mu$m.

Finally, we find our single dust population models to be quite unconstrained as the SEDs of the observed sources need to be represented by mixtures of optically thick and thin dust and ionised gas emission. We discuss qualitative models using multiple, mutually obscuring dust populations in Appendix~\ref{sec:Apx_MultipleDust} while appreciating the caveat that the inclusion of these components further adds to the degeneracy. The derived overall dust masses in these sources for the single dust model are of the order of 10's to 100's $M_{\oplus}$. However, it is important to note that the models with single and multiple dust components do not indicate qualitatively different overall dust masses. The major uncertainties in the dust mass estimates originate from the assumed dust opacities.

\subsection{Pixel-by-pixel spectral index maps}\label{sec:SpIxMap}

\begin{figure*}[htbp]
    \centering
    \includegraphics[width=0.9\textwidth]{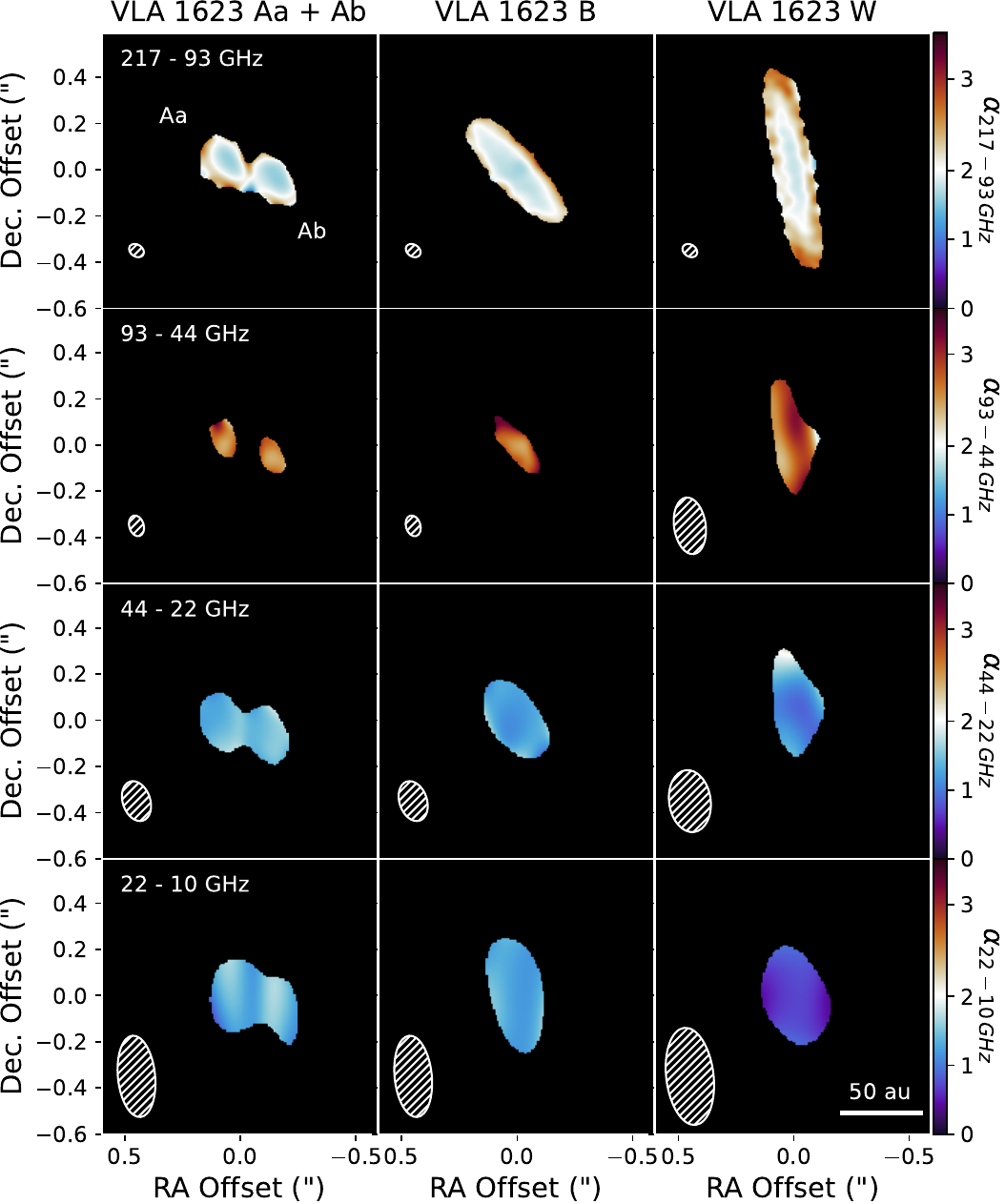}
    \caption{Spectral index maps for each object and each adjacent frequency pair. For B and W, we centre each image on the peak pixel of the object in the highest frequency continuum image and for Aa and Ab we centre the image on the binary midpoint. We show a 50 au scale bar which is shared across all images. We show the common beam for both frequency bands in the bottom left corner as a white ellipse. All images use a 5$\sigma$ mask where $\sigma=\sigma_{rms}$ in Table~\ref{tab:master} except for VLA 1623 W which for 93-44 GHz and 44-22 GHz uses a 3$\sigma$ due to the weak detection at 44 GHz. All images share a common divergent colour bar centred on $\alpha = 2$.}
    \label{fig:SpIx}
\end{figure*}

Further to the disk integrated spectral indices discussed in Sections \ref{sec:SED} and \ref{sec:SED_dustmodel} we are able to probe any spatial variations in the spectral index across individual pixels by comparing images at consecutive frequency pairs for each object. We calculate the spectral index on a per-pixel basis, after smoothing each image to the larger of the beamsizes for each pair of adjacent frequencies in the ALMA and VLA datasets (e.g. from 217--10\,GHz).  We show the results of these calculations in Figure~\ref{fig:SpIx}. 

Examining the $217-93$\,GHz spectral index ($\alpha_{217 - 93\,\text{GHz}}$), we see all objects share a similar spectral index gradient. The general trend of decreasing spectral index from $\sim2.5$ to $\lesssim2$ towards the centre of the emission is shared across all objects likely indicating an optical depth increase caused by high densities near the central protostar. A similar overall trend appears in the 93-44 GHz spectral index ($\alpha_{93 - 44 \text{GHz}}$) for A and B with an increased value of $\sim3$ in the outer regions which decreases to $\sim2.5$ close to the centre. In addition, we find that generally $\alpha_{93 - 44\,\text{GHz}} > \alpha_{217 - 93\,\text{GHz}}$ for all objects indicating a possible reduction in the optical depth at these lower frequencies.

This combination of optically thick emission for $217 - 93\, \text{GHz}$ and more optically thin emission at $93 - 44\,\text{GHz}$ corroborates our SED analysis of the spectral index (see Section \ref{sec:SED}) where we find $\alpha_{\text{Dust}}$ to be just above $\sim$2. This can be explained by multiple populations of optically thick and optically thin dust emission components with mutual obscuration, which we discuss in detail in Appendix \ref{sec:Apx_MultipleDust}.

All objects exhibit much lower spectral indices, typically $\leq2$, for frequency pairs $\leq44\,$GHz. The reduction in spectral index for all objects is correlated with an increased fraction of ionised flux (see Table~\ref{tab:master}) with the lowest spectral indices associated with 22$-$10 GHz frequency pair. Considering this, alongside our spectral index SED analysis in which all objects had $\alpha_{\text{Ionised}}<0.5$, we infer that all objects in this system are subject to ionised gas emission which contributes significantly to observations below 44\,GHz.

VLA~1623~W exhibits a unique morphology compared to the other objects in the system for frequencies $< 93\,$GHz. For $\alpha_{93 - 44\,\text{GHz}}$ we see a decreasing East-West gradient in comparison to the radial gradients seen in A and B. This may be an effect of projection and optical depth whereby the Eastern edge is nearest the observer and thus has a higher column density compared to the western side as mentioned in Section~\ref{sec:JWST_images}.

Additionally, in VLA~1623~W, we see $\alpha_{44 - 22 \text{GHz}} \sim 0.7$ in the very central regions indicating emission from a non-dust component close to the centre of the disk. The spectral index continues to decrease for $\alpha_{22 - 10\text{GHz}}$ which is less than 1 for the entire emission region. Finally, it is worth noting the two lobes of particularly low spectral index ($\sim 0.4$) in the East and West perpendicular to the dust disk. There is also tentative evidence of low spectral index extensions to the West seen in the 93-44\,GHz and 44-22\,GHz spectral index maps.

\subsection{The millimetre continuum emission surface in VLA 1623 B} \label{sec:B_EmSurf}

\begin{figure*}[ht]
    \centering
    \includegraphics[width=\textwidth]{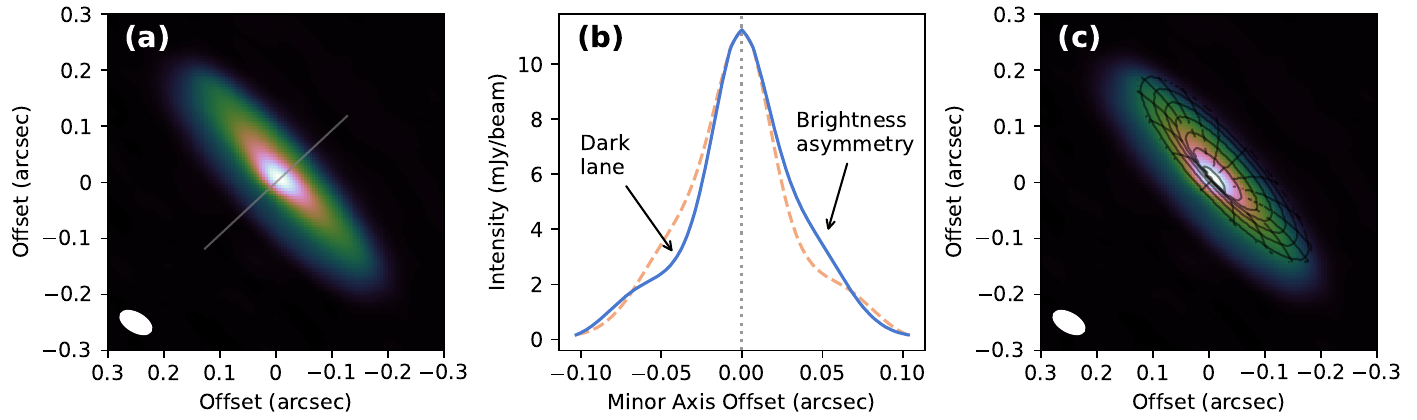}
    \caption{\textbf{(a)} Zoom-in of the 217\,GHz ALMA image of VLA\,1623\,B shown with a colormap to highlight faint features.  The grey line denotes a cut across the minor axis of the disk. \textbf{(b)} Intensity along the minor axis cut from (a) compared with a mirrored profile (dashed line).  There is a dark lane on the SE side and brightness asymmetry on the NW side of the disk detected at a high significance (the rms of 0.02\,mJy\,beam$^{-1}$ is smaller than the line thickness) \textbf{(c)} As (a) but overlaid with a representative surface model at relative heights of $z/r \sim 0.3$ (see text). These features demonstrate the extremely flared nature of the millimetre continuum emission, suggesting that larger grains are vertically well-mixed in the disk.}
    \label{fig:surface}
\end{figure*}

Examining our 217\,GHz observations of VLA\,1623\,B in detail reveals an interesting morphology.  The high resolution of the observations (and the favourable beam orientation along the disk minor axis) means that all of the emission is extremely well resolved. While emission along the disk major axis appears to be symmetric, there is an obvious asymmetry along the minor axis with brighter emission toward the North-West, and a `dark lane' running along the nearside of the disk (see Figure~\ref{fig:surface}). We note that the near- and far-disk sides as seen at 217\,GHz are in agreement with the jet direction shown by SO and SiO emission in \citet{Codella24}, i.e. blue-shifted in the NW direction and redshifted in the SE direction. 

To quantify the vertical extent of the millimetre continuum emission in the VLA\,1623\,B disk, we use the \texttt{GoFish} package \citep{GoFish} to overlay an emission surface that best matches the observed morphology.  This takes the form of a flared disk with an outer taper following the parameterized form
\begin{equation}
    z(r) = z_{0}\left(\frac{r}{1\arcsec}\right)^{\psi}\exp \left(\frac{-r}{r_{\rm taper}} \right)^{q_{\rm taper}},
\end{equation}
where $z_{0}$ is the emission surface at a reference radius, $r$ is the deprojected radius, $r_{\rm taper}$ is the radius of the taper, and the exponents $\psi$, $q_{\rm taper}$ determine the slope of emission surface in the inner and outer disk, respectively.  We experimented with a variety of surface morphologies and found that the observations appear to be well matched by-eye with $z_{0} = 0.5\arcsec$, $\psi = 1.1$, $r_{\rm taper} = 0.23\arcsec$ (32\,au) and $q_{\rm taper}=2.0$, the model of which is shown overlaid on the observations in Figure~\ref{fig:surface}c.  This emission surface varies in terms of the relative height traced throughout the disk, but ranges from $z/r=0.25$--0.35 depending on radius.  This suggests the $\tau=1$ surface for these frequencies in this disk is located high above the midplane, and is further evidence for the optically thick nature of the disk.  In fact, the surface is located at similar heights to the disk regions traced by CO isotopologues observed toward Class II disks \citep{Law2023}, and such a morphological similarity is seen between the gas and dust in other Class 0 disks, for example in the CAMPOS survey \citep{Hsieh24}.  Determining the precise origin of the extremely flared dust morphology requires dedicated radiative transfer modelling of multi-wavelength observations, which we will perform in a future study (Radley et al.\ in prep.).

\section{Discussion}\label{sec:discussion}

\subsection{The unsettled (and unstable?) young disk of VLA~1623~B}\label{sec:unsettled}

Our high resolution observations at 217 and 93 GHz reveal an inclined ($\sim$ 74\degr) disk morphology for VLA~1623~B with significant flaring seen in the 217 GHz continuum that is much less prominent at 93 GHz. In Section \ref{sec:B_EmSurf} we use our 217 GHz observations to determine a $z/r \sim 0.3$ in the inner regions of the disk, and 0.2 in the outer ($>30$ au) regions. Such a morphology is reminiscent of observations of highly inclined Class II protoplanetary disks at visible and infrared wavelengths, in which the brightest emission is the result of scattering of stellar light by dust grains at high relative heights in the disk atmosphere \citep[see, e.g.,][]{Avenhaus2018,Duchene2024}. Similarly high flaring angles and/or dark lane asymmetries have also been observed in other Class 0 and I disks (see e.g.  IRAS~04302+2247, \citealt{Lin23}; R~CrA~IRS7B-a, \citealt{Takakuwa2024ApJ...964...24T}; IRAS4A1, \citealt{Guerra-Alvarado24}). One explanation for these features is that vertical dust settling may occur primarily between the Class I and II stages \citep[see e.g.][]{eDISK} which would explain the more geometrically thin disks found for Class II objects \citep[e.g.][]{Pinte2016}. Alternatively, the suspension of dust grains in the upper disk layers may be due to physical mechanisms operating within young disks.

One such mechanism may be potential gravitational instability operating within the disk of VLA~1623~B. Gravitational instability can trigger spiral shocks in the disk which have been shown to result in significant increases in vertical mixing and turbulence \citep[see e.g.][]{Boley2006}.  Such behaviour would act to lift dust grains that might normally be settled toward the midplane into the upper layers of the disk, resulting in an unsettled disk as observed here.  Class 0 and I protoplanetary disks are deeply embedded, and as such they are expected to be more dynamically unstable than Class II disks due to their increased disk-to-star mass ratio \citep{Vorobyov06,Tsukamoto17}. To accurately calculate the disk-to-star mass ratio we require high resolution gas observations of the inner regions of protoplanetary disks. Recent radiative transfer models focusing on C$^{17}$O kinematic observations have constrained the protostellar mass of VLA~1623~B to be 1.9 $M_{\odot}$ \citep{Sadavoy24}. This tracer is less subject to contamination from infalling envelopes or streamers and thus provides a robust estimate of the stellar mass.

A disk-to-star mass ratio $\ge0.1$ indicates that the disk's self gravity may be important and thus influence further evolution of the system \citep{Kratter2016}. If we use the disk dust mass of 650$M_{\oplus}$ derived in Section \ref{sec:SED_dustmodel} and a standard 1:100 dust-to-gas mass ratio, we estimate a disk-to-star mass ratio of $\sim$\,0.1 which implies that the disk may be gravitationally unstable. It is important to note that our observations at 217\,GHz and potentially at 93\,GHz are optically thick due to high densities and a high inclination. This means we are only tracing the $\tau_\nu \sim 1$ emission surface \citep{Delussu24,Chung2024ApJS..273...29C} and so our disk mass and therefore disk-to-star mass ratio is likely to be a lower limit. With this consideration, we believe it is very likely that the disk is gravitationally unstable.

Alternatively, the lack of settling may arise from infalling material which can significantly perturb the disk and excite instabilities such as spiral waves entraining dust grains into spiral arms \citep[e.g.][]{Bae2015}. VLA~1623~B has been observed to have wide line widths in SO \citep[$>10\text{\,km\,s}^{-1}$,][]{Hsieh20} believed to originate from shocks caused by infalling material. Additionally, in the presence of anisotropic infall, the models of \citet{Kuznetsova22} find a mean dust disk diameter of $\sim55$\,au which is slightly larger than the dust disk we derive for B at 217\,GHz ($\sim40$\,au). However, since this model focuses on higher-mass cores, we would expect this to be lower in low-mass objects such as VLA~1623~B. Therefore, infall may have a significant impact on the circumstellar disk of VLA~1623~B.

Nevertheless, if the disk of VLA~1623~B is unstable, due to gravitational instability or infall, then we would expect the manifestation of axisymmetric structures such as spiral density waves \citep{Mayer04,Bae2015,Huang18} which are not observed.  However, as the disk is optically thick at 217\,GHz we are not tracing through all layers of emission and thus any existing substructures may remain hidden \citep[see e.g.][]{Xu2022}. Alternatively, \citet{Tsukamoto17} finds that spirals induced by GI may be short lived and only persist for $10\%$ of the Class 0/I disk's lifetime. If the disk contains non-axisymmetric structures, we would require further observations at a more optically thin frequency that matches or improves upon the resolution of our 217\,GHz observation.  With currently available instrumentation, this would present a challenge.

\subsection{Environmental impact on disk evolution}

\citet{Hsieh24} performed an ALMA Band 6 (217\,GHz) survey probing the effect of evolutionary state and star forming environment on the observed protostellar disk radii. Using similarly high resolution observations ($\sim0\farcs1$), they found dust disks in Ophiuchus to have an average radii of 35\,au and 26\,au for Class 0 and I objects, respectively, meaning that Ophiuchus YSOs have a smaller average dust disk radii than Serpens (77\,au and 49\,au) and Aquila (50\,au and 62\,au) for example \citep{Hsieh24}. We find that VLA~1623~W has a circumstellar dust disk radius of $40.3\pm 0.4$\,au which aligns with previous 217\,GHz observations \citep[e.g. 48\,au,][]{Mercimek23}. This radius is therefore just under double the average Class I dust disk radius of Ophiuchus as a whole and significantly larger than the other members of the VLA~1623 system. We see that VLA~1623~B is roughly half the size of VLA~1623~W at $19.4\pm0.1$\,au corroborating previous estimations of the dust disk size at 217\,GHz (see e.g. 23\,au \citealt{Harris18}; 22\,au \citealt{Mercimek23}). However, as mentioned in section \ref{sec:unsettled} the small size of VLA~1623~B may be due to the influence of infalling material perturbing the disk evolution \citep[e.g.][]{Kuznetsova22}. 

The most extreme difference is seen in the VLA~1623~Aa and Ab circumstellar dust disks which have radii of $7.37\pm0.04$\,au and $6.81\pm0.04$\,au, respectively, making them roughly one fifth the size of the average Class 0 population in Ophiuchus. However, small protostellar dust disks are not uncommon for binary systems \citep[see e.g.][]{Long19,Manara19, Narayanan23} and may be caused by disk truncation through binary interaction if Aa and Ab are close enough to interact with each other \citep[see e.g.][]{Artymowicz94}.

It appears that in comparison to the overall population of Ophiuchus, VLA~1623~Aa, Ab, and B have significantly smaller radial extents whereas VLA~1623~W has a significantly larger radial extent. Recent ALMA-FAUST images of C$^{18}$O(2–1) \citep{Mercimek23} revealed streamers that extend up to 1000 au, spatially and kinematically connecting the envelope and outflow cavities of the A1 + A2 + B system with the disc of VLA~1623~W. The presence of the streamers, as well as the spatial (1300 au) and velocity (2.2 km s$^{-1}$) offsets of VLA~1623~W, suggests two main scenarios:  either the sources W and A + B originated in separate cores and have interacted with each other \citep[as suggested by][]{Sadavoy24}, or source W was expelled from the VLA~1623 multiple system while it was forming.
In the latter case, the streamers could channel material from the envelope and cavities of VLA 1623 A + B towards VLA 1623 W, thereby contributing to the determination of its ultimate mass and chemical composition \citep{Mercimek23}.

\subsection{Evidence for jet emission}

Protostellar objects are often associated with jets, outflows and disk winds \citep[see e.g.][]{Macias16}. Winds and jets can exhibit very similar spectral indices and thus require further investigation, often using morphological arguments to discern which mechanism is in operation \citep{Anglada18}.  Here we examine whether our VLA observations reveal the presence of any such phenomena in the members of VLA~1623.

\subsubsection{VLA 1623AB}

VLA~1623~Aa and Ab have both been associated with their own molecular outflows \cite[e.g.][]{Hara21} which necessitates a launching mechanism such as a protostellar jet  or a disk wind \citep[see][for reviews]{Ray21, Pascucci22}. Through our SED analysis we find that both objects exhibit a significant ionised gas contribution at low frequencies ($>50\%$ at 10\,GHz). Considering the upper limit caveats at low frequencies for Aa and Ab, we note that both objects have a integrated spectral index consistent with a thermal radio jet \citep{Anglada18}.

We find an indication of jet-like behaviour for Aa at 10\,GHz where we see a sudden increase in the flux and size of Aa compared to Ab which is accompanied by an extension with a position angle of 110$\,\pm\, 20\degr$. The extension is perpendicular to the position angle of the dust disk derived at 217\,GHz, 30.9$\,\pm\,0.5\degr$, as would be naively expected from jet emission \citep{Anglada18}. If the emission is arising from a jet or a wind then the emission region must be at sub-beam scales ($<20$ au) which supports either the disk launching theory proposed in \citet{Ohashi22} at a radius of $\sim5-16$ au or is evidence of a protostellar jet.

Further to this, using CS observations, \citet{Ohashi22} traced the origin of the outflow associated with the VLA 1623 A binary to a position close to Aa but could not definitively determine the true launch site due to low spatial resolution. Additionally, \citet{Hsieh24} postulates that the misalignment of the Aa and Ab outflows detected in \citet{Hara21} could be explained by an outflow launched by B which is misaligned to the VLA 1623 A binary. 

In Figure~\ref{fig:F444W}b we see an outflow cavity that is cospatial with the 217\,GHz continuum of VLA~1623~B. As discussed in Section~\ref{sec:JWST_images} this outflow cavity is aligned with the expected position angle of a highly collimated jet predicted by \citet{Santangelo2015}. Unlike VLA~1623~Aa we do not detect a perpendicular emission component at 10\,GHz. However, as the jet is expected to be highly collimated, which may also explain the smaller opening angle (75\degr) compared to Aa (135\degr), the launching region may also occupy a much smaller spatial extent. 

The proximity of the conical emission in our JWST NIRcam data (see Figure~\ref{fig:F444W}a and \ref{fig:F444W}b) to Aa and B point towards both objects driving individual outflows. This result confirms the previous hypotheses of \citet{Hsieh20} and \citet{Ohashi22}. Furthermore, due to the increased flux and perpendicular PA at 10\,GHz for Aa, we believe these observations suggests that Aa is more likely to be the origin of the large scale outflow.

There are two possible solutions to improve our understanding of these objects. Firstly, we can use time-series observations to probe variations in the 10\,GHz flux, which we attribute primarily to changes in the contribution of ionised gas.  Secondly, we may benefit from increased integration times and therefore sensitivities to pick up on any fainter emission that may arise if Ab has a different jet properties to Aa.

\subsubsection{VLA 1623 W}

VLA 1623 W has not been associated with a jet or an outflow in previous studies, possibly due to its low flux and the previous ambiguity of its nature \citep[e.g.][]{Mercimek23,Hara21}. In contrast, through our SED fitting we find a negative integrated spectral index at low frequencies ($< 10\,$GHz), which is not seen in any other object in the VLA~1623 system. We also see a general trend of decreasing spectral index with decreasing frequency in both the integrated and pixel-by-pixel spectral indices as well as an increased contamination fraction at 10 and 22\,GHz. This indicates significant ionised gas emission dominates at lower frequencies, which is similarly found in other Class I disks \cite[see e.g. IRAS~04302+2247;][]{Villenave23}.

At 44\,GHz we see evidence of a tenuous ($\sim2.5-3\sigma$) extension perpendicular to the plane of the disk which coincides with a low $\alpha_{93-44\text{GHz}}$ value in the spectral index map. We note that this extension is at the noise level and the spectral index map may be subject to edge effects. However, at 10\,GHz and 22\,GHz we see similar extensions (although not perfectly aligned) at both the 3- and 5-$\sigma$ level. Fortuitously, these extensions are also oriented such that we have the best spatial resolution in that direction. The $\alpha_{22-10\text{GHz}}$ spectral index map also shows two lobes of reduced spectral index ($\sim0.5$) in a similar direction to the extensions seen in the continuum. The combination of a low spectral index as well as perpendicular elongations may indicate the existence of a jet in VLA~1623~W. To confirm this, we would require higher resolution and sensitivity observations to pin point the origin of the outflow as well as any tenuous emission associated with it.

\section{Conclusions}\label{sec:conclusion}

In this paper we have investigated the protostellar system VLA~1623 using JWST, ALMA and VLA.  The high spatial resolution across each of the instruments (4-30\,au) reveals unprecedented detail in one of the nearest young protostellar systems across nearly four orders of magnitude in wavelength.  Below we summarise our main findings.      

\begin{itemize}

\item Using JWST 4.44\,\micron\ observations we detect two outflow cavities in scattered light with origins that are cospatial with ALMA and VLA continuum observations of VLA~1623~Aa and B. The conical morphology and orientation of the emission suggests that these observations trace outflowing material from Aa and B. The proximity of the outflow cavity to Aa in addition to the flux increase and perpendicular morphology at 10\,GHz implies this object is responsible for the large scale outflow seen in VLA~1623.

\item VLA~1623~B appears to be a highly inclined disk at 217 and 93\,GHz, with an optical depth surface for the former originating from $z/r \sim 0.3$, implying significant dust settling has yet to take place in the disk around this young Class 0 object. Based on the disk-to-star mass ratio of VLA~1623~B, it is possible that gravitational instabilities may be operating in the disk, which would naturally lead to an increased dust scale height.

\item Fitting the mm-cm SEDs for each object reveals that significant dust growth ($a_{\text{max}}\gtrsim1$\,mm) may have already taken place at these early stages of disk evolution. Furthermore, all objects appear to be optically thick at ALMA frequencies (217\,GHz and 93\,GHz) and dominated by ionised gas emission at low ($\lesssim 15$\,GHz) frequencies. For VLA~1623~B and W, the total dust mass is high ($\sim$100's $M_{\oplus}$) and likely a lower limit due to the aforementioned high optical depths.

\item For frequencies below 44\,GHz, we note non-negligible contributions to the flux from the ionised gas components (e.g. $\gtrsim$ 5--20 per cent).  This suggests frequencies between 44 and 93\,GHz may offer the best trade-off between detecting optically thin dust emission while ensuring minimal contamination from ionised gas in YSOs, although this depends on the properties of the individual objects.

\end{itemize}

Our results represent some of the highest resolution images currently available across these frequency regimes, particularly in the case of the A-configuration VLA observations.  Despite this, much of our analysis is limited by the fact that we are unable to accurately resolve the emission morphology at frequencies below 22\,GHz.  Higher resolution observations will only be available with next-generation interferometers operating at these frequencies, such as the Square Kilometre Array (SKA) and next-generation Very Large Array (ngVLA).  At full capabilities, SKA-Mid will achieve angular resolutions approximately 6 times higher at 10\,GHz than the observations we present here \citep[55 mas, see][]{Braun2019}.  The ngVLA will further increase this resolution by a factor of $\sim7$ at the same frequencies \citep{Selina2018}.  Such a leap in capability will be comparable to (or even better than) the highest resolution infrared and (sub-)millimetre observations currently available, opening the door to truly resolved studies of YSOs across several orders of magnitude in frequency for the first time.

\section*{Acknowledgements}

We thank the referee for a constructive report which helped to improve the clarity of the manuscript.  
ICR acknowledges a studentship funded by the Science and Technology Facilities Council of the United Kingdom (STFC).
GB is supported by a 2022 Leonardo Grant for Researchers in Physics from the BBVA Foundation. The BBVA Foundation takes no responsibility for the opinions, statements and contents of this project, which are entirely the responsibility of its authors. GB aned JMG acknowledges support from the PID2020-117710GB-I00 grant funded by MCIN/AEI/ 10.13039/501100011033 and from the PID2023-146675NB-I00 (MCI-AEI-FEDER, UE) program. 
JDI acknowledges support from an STFC Ernest Rutherford Fellowship (ST/W004119/1) and a University Academic Fellowship from the University of Leeds. 
H.B.L. is supported by the National Science and Technology Council (NSTC) of Taiwan (Grant Nos.\ 111-2112-M-110-022-MY3, 113-2112-M-110-022-MY3).
CC, and LP acknowledge the EC H2020 research and innovation programme for the project "Astro-Chemical Origins” (ACO, No 811312), the PRIN-MUR 2020  BEYOND-2p (Astrochemistry beyond the second period elements, Prot. 2020AFB3FX), the project ASI-Astrobiologia 2023 MIGLIORA (Modeling Chemical Complexity, F83C23000800005), the INAF-GO 2023 fundings PROTO-SKA (Exploiting ALMA data to study planet forming disks: preparing the advent of SKA, C13C23000770005), the INAF Mini-Grant 2022 “Chemical Origins”. 
LP, ClCo, and E.B. also acknowledge financial support under the National Recovery and Resilience Plan (NRRP), Mission 4, Component 2, Investment 1.1, Call for tender No. 104 published on 2.2.2022 by the Italian Ministry of University and Research (MUR), funded by the European Union – NextGenerationEU– Project Title 2022JC2Y93 Chemical Origins: linking the fossil composition of the Solar System with the chemistry of protoplanetary disks – CUP J53D23001600006 - Grant Assignment Decree No.\ 962 adopted on 30.06.2023 by the Italian Ministry of Ministry of 
University and Research (MUR). 
E.B. acknowledges support from Next Generation EU funds within the National Recovery and Resilience Plan (PNRR), Mission 4 - Education and Research, Component 2 - From Research to Business (M4C2), Investment Line 3.1 - Strengthening and creation of Research Infrastructures, Project IR0000034 – “STILES - Strengthening the Italian Leadership in ELT and SKA”.
L.L. acknowledges the support of DGAPA PAPIIT grants IN108324 and IN112820 and CONACyT-CF grant 263356.
I.J-.S acknowledges funding from grant No. PID2022-136814NB-I00 funded by the Spanish MICIU/AEI/ 10.13039/501100011033 and by “ERDF/EU”.

We acknowledge the Spanish Prototype of an SRC (SPSRC) service and support funded by the Spanish Ministry of Science, Innovation and Universities, by the Regional Government of Andalusia, by the European Regional Development Funds and by the European Union NextGenerationEU/PRTR. The SPSRC acknowledges financial support from the State Agency for Research of the Spanish MCIU through the "Center of Excellence Severo Ochoa" award to the Instituto de Astrof\'isica de Andaluc\'ia (SEV-2017-0709) and from the grant CEX2021-001131-S funded by MCIN/AEI/ 10.13039/501100011033.
A portion of this research was carried out at the Jet Propulsion Laboratory, California Institute of Technology, under a contract with the National Aeronautics and Space Administration (80NM0018D0004).
This work is based in part on observations made with the NASA/ESA/CSA James Webb Space Telescope. The data were obtained from the Mikulski Archive for Space Telescopes at the Space Telescope Science Institute, which is operated by the Association of Universities for Research in Astronomy, Inc., under NASA contract NAS 5-03127 for JWST. These observations are associated with program \#2739 and can be accessed via \dataset[DOI]{10.17909/js8e-9q91}.
This paper makes use of the following ALMA data: ADS/JAO.ALMA\#2019.1.01074.S ALMA is a partnership of ESO (representing its member states), NSF (USA) and NINS (Japan), together with NRC (Canada), MOST and ASIAA (Taiwan), and KASI (Republic of Korea), in cooperation with the Republic of Chile. The Joint ALMA Observatory is operated by ESO, AUI/NRAO and NAOJ. The National Radio Astronomy Observatory is a facility of the National Science Foundation operated under cooperative agreement by Associated Universities, Inc.

\facilities{JWST (NIRCam), ALMA, VLA}

\software{CASA \citep{CASA}, analysisUtils \citep{analysisutils_2023}, JWST Calibration Pipeline \citep{Bushouse23}, Astropy \citep{astropy2013,astropy2018,astropy2022}, SciPy \citep{SciPy}, PyBDSF \citep{PyBDSF}, emcee \citep{Emcee}, GoFish \citep{GoFish}.}

\bibliography{Bibliography.bib}

\appendix
\label{appendix}
\restartappendixnumbering

\section{Observational details}

\begin{deluxetable*}{lccccc}[h]
\tablecaption{Details of the ALMA and VLA observations and their measured properties for the variability and fiducial image products. \label{tab:obs_params}}
\tabletypesize{\footnotesize}
\tablehead{\colhead{Band} & \colhead{6} & \colhead{3} & \colhead{Q} & \colhead{K} & \colhead{X}}
\startdata
Central wavelength  &   1.4\,mm & 3.2\,mm   & 7.0\,mm   & 1.4\,cm   & 3.0\,cm   \\
Central frequency   &    217\,GHz       &     93\,GHz      &    44\,GHz        &    22\,GHz         &    10\,GHz       \\     
Observing date (UT) & 2021 Aug 7; 2021 Oct 3, 7 & 2021 Sep 7 & 2022 Apr 2, 7 & 2022 Apr 2, 7 & 2022 Apr 2, 7\\
Project code   & 2019.1.01074.S & 2019.1.01074.S & 22A-164 & 22A-164 & 22A-164\\
Configuration & C43-8 & C43-9/10 & A & A & A \\
Phase Center (J2000): \\
\phantom{x} R.A. &\phantom{$-$}16:26:26.42 &\phantom{$-$}16:26:26.42 &\phantom{$-$}16:26:27.80 &\phantom{$-$}16:26:25.63 &\phantom{$-$}16:26:24.60 \\
\phantom{x} Dec. &$-$24.24.30.00
&$-$24.24.30.00 &$-$24.24.40.30 &$-$24.24.29.4 &$-$24.23.37.00 \\
Primary beam size (FWHP) & 27\arcsec & 62\arcsec & 1.0\arcmin & 2.0\arcmin & 4.5\arcmin \\
Maximum recoverable scale & 0\farcs85 & 0\farcs86 & 1\farcs2 & 2\farcs4 & 5\farcs3 \\
Total time on source (mins) & 86.8& 84.5& 60.7&28.7&5.5\\
Full bandwidth (GHz) &5.6&7.5&7.9 &7.9 &4\\
Gain calibrator     & J1633-2557 & J1633-2557 & J1625-2527 & J1625-2527 & J1625-2527\\
Bandpass calibrator & J1337-1257 & J1517-2422 & J1256-0547 & J1256-0547 & J1256-0547 \\
Flux calibrator     & J1337-1257 & J1517-2422 &3C286 &3C286 &3C286\\
& \\
%
\hline
\multicolumn{6}{c}{\textbf{Variability Image properties:}}\\
\sidehead{2 April 2022}
Briggs Robust = 2.0 & & & & & \\
\phantom{x} Synthesized beam & \nodata & \nodata & 0\farcs11$\times$0\farcs1 [28\degr]  & 0\farcs26$\times$0\farcs14 [26\degr] & 0\farcs43$\times$0\farcs19 [7\degr]\\
\phantom{x} Cont. rms (mJy\,beam$^{-1}$)\tablenotemark{a} & \nodata & \nodata & 0.03 & 0.01 & 0.03\\
\sidehead{7 April 2022}
Briggs Robust = 2.0 & & & & & \\
\phantom{x} Synthesized beam & \nodata & \nodata & 0\farcs11$\times$0\farcs08 [8\degr] &  0\farcs2$\times$0\farcs18 [17\degr]  & 0\farcs42$\times$0\farcs22 [6\degr]\\
\phantom{x} Cont. rms (mJy\,beam$^{-1}$)\tablenotemark{a} & \nodata & \nodata & 0.03 & 0.01 & 0.02\\
& \\
\hline
\multicolumn{6}{c}{\textbf{Fiducial Image properties:}} \\
Briggs Robust = 0 & & & & & \\
\phantom{x} Synthesized beam & 0\farcs06$\times$0\farcs03 [60\degr] & 0\farcs07$\times$0\farcs05 [57\degr] & \nodata & \nodata & \nodata \\
\phantom{x} Cont. rms (mJy\,beam$^{-1}$)\tablenotemark{a} & 0.02 & 0.01 & \nodata & \nodata & \nodata \\
Briggs Robust = 0.5 & & & & & \\
\phantom{x} Synthesized beam & \nodata & \nodata & 0\farcs09$\times$0\farcs06 [17\degr] & 0\farcs18$\times$0\farcs12 [19\degr] & 0\farcs35$\times$0\farcs16 [6\degr] \\
\phantom{x} Cont. rms (mJy\,beam$^{-1}$)\tablenotemark{a} & \nodata & \nodata & 0.02 & 0.009 & 0.01\\
Briggs Robust = 2.0 & & & & & \\
\phantom{x} Synthesized beam & \nodata & \nodata & \nodata & 0\farcs22$\times$0\farcs16 [24\degr] & 0\farcs42$\times$0\farcs21 [7\degr]\\
\phantom{x} Cont. rms (mJy\,beam$^{-1}$)\tablenotemark{a} & \nodata & \nodata & \nodata & 0.01 & 0.01\\
UVTaper = 0\farcs2$\times$0\farcs08 [15\degr] & & & & & \\
\phantom{x} Synthesized beam & \nodata & \nodata & 0\farcs26$\times$0\farcs14 [8\degr] & \nodata & \nodata\\
\phantom{x} Cont. rms (mJy\,beam$^{-1}$)\tablenotemark{a} & \nodata & \nodata & 0.03 & \nodata & \nodata\\
\enddata
\tablenotetext{a}{Measured at the centre of the primary beam.}
\end{deluxetable*}

\clearpage
\section{Variability between VLA epochs}

\begin{deluxetable*}{lcccccc}[h]
\tablecaption{Flux comparison of VLA observations at each epoch for a robust weighting of 2. \label{tab:EpochFluxes}}
\tablehead{\colhead{Object} & \colhead{Frequency (GHz)} & \colhead{F$_{min}$ (mJy)} & \colhead{F$_{max}$ (mJy)} & \colhead{$\Delta$ F(mJy)} & \colhead{$\Delta$F error (mJy)} & \colhead{Variable?}}
\startdata
VLA1623B & 10 & 0.37 & 0.41 & 0.04 & 0.08 & N \\
 & 22 & 0.98 & 1.09 & 0.11 & 0.15 & N \\
 & 44 & 2.51 & 2.62 & 0.11 & 0.43 & N \\
 \\
 \hline
VLA1623Ab & \phantom{x}10\tablenotemark{a} & 0.41 & 0.52 & 0.11 & 0.13 & N \\
 & 22& 0.41 & 0.50 & 0.09 & 0.08 & N \\
 & 44 & 1.29 & 1.40 & 0.12 & 0.25 & N \\
 \\
 \hline
VLA1623Aa & \phantom{x}10\tablenotemark{a} & 0.41 & 0.52 & 0.11 & 0.13 & N \\
 & 22 & 0.46 & 0.46 & 0.00 & 0.07 & N \\
 & 44 & 1.34 & 1.48 & 0.14 & 0.26 & N \\
 \\
 \hline
VLA1623W & 10 & 0.25 & 0.25 & 0.04$\times 10^{-1}$ & 0.07 & N \\
 & 22 & 0.35 & 0.37 & 0.02 & 0.07 & N \\
 & \phantom{x}44\tablenotemark{b}& \nodata& \nodata& \nodata& \nodata&\nodata
 \\
 \\
\enddata
\tablenotetext{a}{VLA1623Aa and Ab are unresolved at X-band for a robust weighting 2 and thus share the same flux measurements.}
\tablenotetext{b}{Image fidelity for each epoch makes accurate flux measurements unreliable to calculate.}
\end{deluxetable*}

\clearpage
\section{SED fluxes}

\begin{deluxetable*}{lccccc}[h]
\tablecaption{Flux measurements for each object used in Figures~\ref{fig:SED} and \ref{fig:SingPopSED}. \label{tab:Fluxes}}
\tablehead{\colhead{Object} & \colhead{Band} & \colhead{Frequency (GHz)} & \colhead{Robust} & \colhead{Total Flux (Jy)} & \colhead{Total Flux error (Jy)} }
\startdata
VLA~1623~Aa & \phantom{x}C\tablenotemark{$\dagger$} & 4.5& \nodata & $8.70\times 10^{-5}$ & $3.03\times 10^{-5}$ \\
 & \phantom{x}C\tablenotemark{$\dagger$} & 7.5& \nodata & $1.25\times 10^{-4}$ & $2.57\times 10^{-5}$ \\
 & X & 10 & 0.5 & $2.33\times 10^{-4}$ & $4.61\times 10^{-5}$ \\
 & K & 22 & 0.5 & $4.62\times 10^{-4}$ & $5.00\times 10^{-5}$ \\
 & Q & 44 & 0.5 & $1.25\times 10^{-3}$ & $1.56\times 10^{-4}$ \\
 & B3 & 93 & 0.0 & $9.00\times 10^{-3}$ & $4.50\times 10^{-4}$ \\
 & B6 & 217 & 0.0 & $4.54\times 10^{-2}$ & $4.54\times 10^{-3}$ \\
 & \phantom{x}B7\tablenotemark{$\ast$} & 350& \nodata & $1.93\times 10^{-1}$ & $2.17\times 10^{-2}$ \\
  \hline
VLA~1623~Ab & \phantom{x}C\tablenotemark{$\dagger$} & 4.5& \nodata & $8.70\times 10^{-5}$ & $3.03\times 10^{-5}$ \\
  & \phantom{x}C\tablenotemark{$\dagger$} &  7.5& \nodata & $1.25\times 10^{-4}$ & $2.57\times 10^{-5}$ \\
  & X &  10 & 0.5 & $1.16\times 10^{-4}$ & $3.27\times 10^{-5}$ \\
  & K &  22 & 0.5 & $4.40\times 10^{-4}$ & $4.83\times 10^{-5}$ \\
  & Q &  44 & 0.5 & $1.24\times 10^{-3}$ & $1.52\times 10^{-4}$ \\
  & B3 &  93 & 0.0 & $8.76\times 10^{-3}$ & $4.38\times 10^{-4}$ \\
  & B6 &  217 & 0.0 & $3.97\times 10^{-2}$ & $3.97\times 10^{-3}$ \\
  & \phantom{x}B7\tablenotemark{$\ast$} &  350& \nodata & $1.52\times 10^{-1}$ & $1.63\times 10^{-2}$ \\
  \hline
VLA~1623~B & \phantom{x}C\tablenotemark{$\dagger$} & 4.5& \nodata & $1.89\times 10^{-4}$ & $3.52\times 10^{-5}$ \\
  & \phantom{x}C\tablenotemark{$\dagger$} &  7.5& \nodata & $1.89\times 10^{-4}$ & $3.52\times 10^{-5}$ \\
  & X &  10 & 0.5 & $3.55\times 10^{-4}$ & $3.96\times 10^{-5}$ \\
  & K &  22 & 0.5 & $1.03\times 10^{-3}$ & $1.09\times 10^{-4}$ \\
  & Q &  44 & 0.5 & $2.45\times 10^{-3}$ & $3.09\times 10^{-4}$ \\
  & B3 &  93 & 0.0 & $2.16\times 10^{-2}$ & $1.08\times 10^{-3}$ \\
  & B6 &  217 & 0.0 & $1.07\times 10^{-1}$ & $1.07\times 10^{-2}$ \\
  & \phantom{x}B7\tablenotemark{$\ast$} &  350& \nodata & $3.21\times 10^{-1}$ & $3.25\times 10^{-2}$ \\
  \hline
VLA~1623~W & \phantom{x}C\tablenotemark{$\dagger$} & 4.5& \nodata & $2.18\times 10^{-4}$ & $1.78\times 10^{-5}$ \\
  & \phantom{x}C\tablenotemark{$\dagger$} &  7.5& \nodata & $1.98\times 10^{-4}$ & $2.51\times 10^{-5}$ \\
  & X &  10 & 2.0 & $2.42\times 10^{-4}$ & $3.84\times 10^{-5}$ \\
  & K &  22 & 2.0 & $3.68\times 10^{-4}$ & $4.62\times 10^{-5}$ \\
  & Q &  44 & 2.0 & $4.90\times 10^{-4}$ & $1.36\times 10^{-4}$ \\
  & B3 &  93 & 0.0 & $8.20\times 10^{-3}$ & $4.10\times 10^{-4}$ \\
  & B6 &  217 & 0.0 & $5.18\times 10^{-2}$ & $5.18\times 10^{-3}$ \\
  & \phantom{x}B7\tablenotemark{$\ast$} &  350& \nodata & $1.59\times 10^{-1}$ & $1.67\times 10^{-2}$ \\
\enddata
\tablenotetext{\dagger}{Taken from \citet{Dzib13}}
\tablenotetext{\ast}{Taken from \citet{Harris18}}
\end{deluxetable*}

\clearpage
\section{Astrometric Alignment}\label{sec:APXAstrometry}
\begin{figure*}[htbp]
    \centering
    \includegraphics[width=\textwidth]{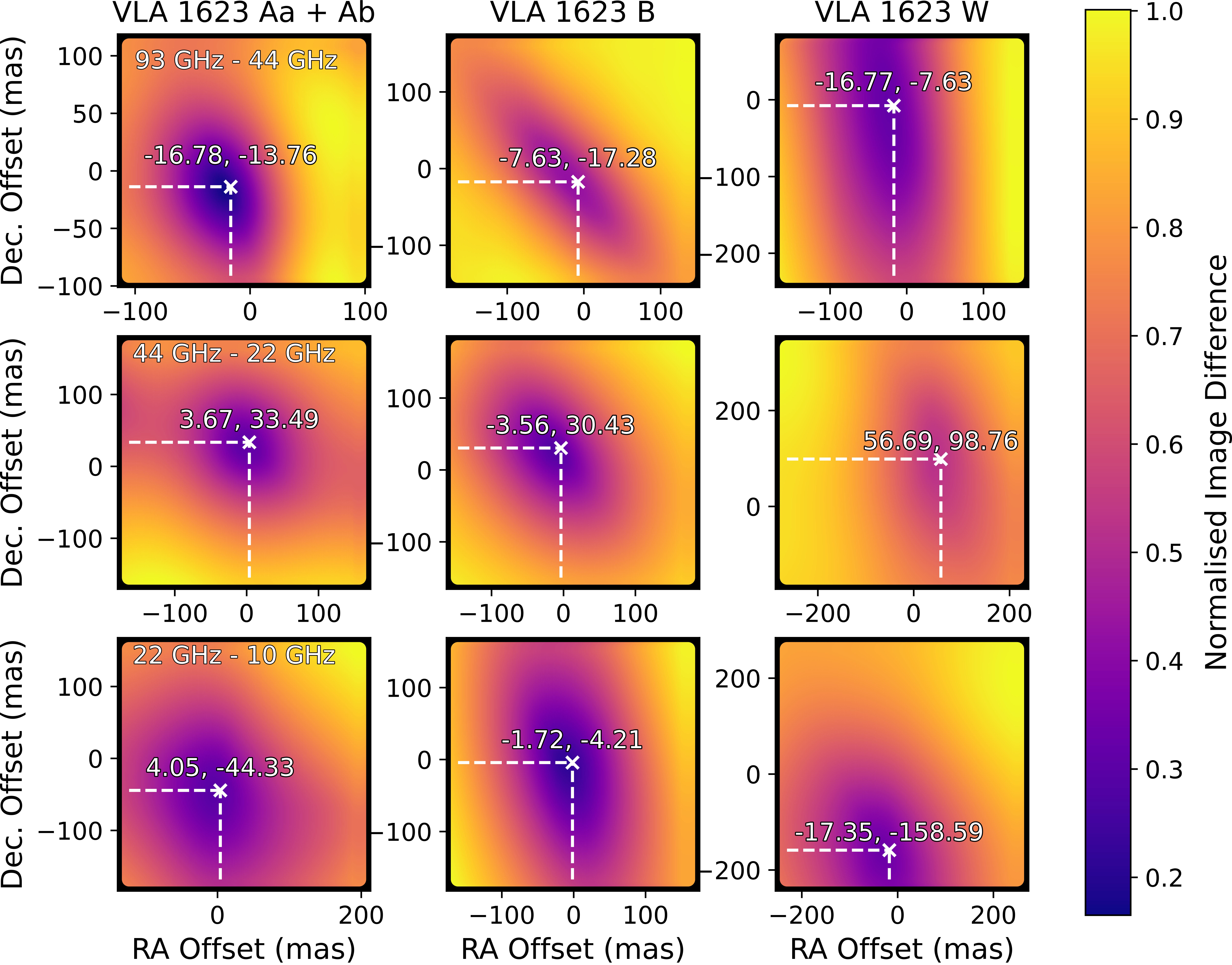}
    \caption{Following the methodology in Section \ref{sec:astrometric} we present derived image shifts for each set of images used in Section \ref{sec:SpIxMap}. The colour scale represents our image difference metric, normalised by the maximum image difference value in each object and frequency pair considered. We indicate the corresponding spectral index pair for each row in the top left of the first column. Each column is titled with the object(s) for which the minimisation was performed. We indicate the optimal image shift as a white cross annotated with the value of the offsets.}
    \label{fig:MinimisationLandscape}
\end{figure*}

In Figure~\ref{fig:MinimisationLandscape} we present the resulting image shifts from our image difference minimisation discussed in Section \ref{sec:astrometric}. These shifts are used to overlay images for each frequency pair used in our spectral index map analysis (see Section \ref{sec:SpIxMap}). We quantify the difference between two images after each shift by considering the absolute value of the difference in pixel values i.e. $|x_1-x_2|$ where $x_1$ and $x_2$ are overlapping pixels from image 1 and image 2, respectively. Using this image difference metric, we normalise the value by the maximum difference from the shifts considered and refer to this as the normalised image difference, presented as the colour scale in Figure~\ref{fig:MinimisationLandscape}.

\clearpage
 \section{SED Modelling with multiple dust populations}

 \subsection{Ionised gas} emission\label{appendix:freefree}

Following \citet{Keto2003ApJ...599.1196} and \citet{Mezger1967ApJ...147..471M}, we approximate $\tau_{\nu}^{\text{Ionised}}$ by:
\begin{equation}
\label{eq:tauff}
\tau_{\nu}^{\text{Ionised}}=8.235\times10^{-2}\left(\frac{T_{e}}{\mbox{K}}\right)^{-1.35}\left(\frac{\nu}{\mbox{GHz}}\right)^{-2.1}\left(\frac{\mbox{EM}}{\mbox{pc\,cm$^{-6}$}}\right),
\end{equation}
where $T_{e}$ is the electron temperature, and EM is the emission measure defined as EM$=$$\int n_{e}^{2}d\ell$, $n_{e}$ is the electron number volume density.
The flux density $F_{\nu}$ is given by
\begin{equation}
\label{eq:dust}
F_{\nu} =B_{\nu}(T_{e})(1-{e}^{-\tau_{\nu}^{\text{Ionised}}})\Omega_{\text{Ionised}},
\end{equation}
where $\Omega_{\text{Ionised}}$ is the observed solid angle, and $B_{\nu}(T)$$=$$(2h\nu^{3}/c^{2})$$(e^{h\nu/k_{B}T} -1)^{-1}$ is the Planck function at temperature $T_{e}$, $h$ and $k_{B}$ are Planck and Boltzmann constants. 

The ionised gas emission is optically thick at low frequencies, with spectral index $\alpha$ as high as 2.0; it is optically thin at high frequencies where $\alpha$ can be as low as $-$0.1. Given $T_{e}$, the turn-over frequency that separates the optically thick and thin regime is determined by EM.

\subsection{Multiple dust populations} \label{sec:Apx_MultipleDust}
In order to more accurately describe the SEDs of each source in Section~\ref{sec:SED_dustmodel}, we must use a minimum of 10 free parameters i.e. $T_{\text{dust}}$, $\Sigma_{\text{dust}}$, $a_{\text{max}}$ and $\Omega_{\text{dust}}$ for two dust components, as well as EM and $\Omega_{\text{Ionised}}$ for the ionised gas component.  Moreover, we found that the SEDs of our target sources are not necessarily consistent with uniform $T_{\text{dust}}$, $\Sigma_{\text{dust}}$, $a_{\text{max}}$.

To explain the strong emission at centimetre wavelengths, we needed to assume the geometry that dust components with higher $T_{\text{dust}}$  are obscured by dust components with lower $T_{\text{dust}}$ in the line of sight, which is an assumption that is often needed in the modelling of the (sub)millimetre SEDs of deeply embedded Class 0/I objects (c.f., \citealt{Li2017ApJ...840...72L}). In addition, we found that the higher $T_{\text{dust}}$ components in a source may have higher $a_{\text{max}}$ and/or higher $\Sigma_{\text{dust}}$ than the lower $T_{\text{dust}}$ components, although the values of $\Sigma_{\text{dust}}$ and $a_{\text{max}}$ are degenerate in our models. As a consequence, we require more free parameters to describe the SEDs of each object. However, since our SEDs include at most eight independent measurements for each source, the  parameters for our SED models are inevitably degenerate. Therefore, we do not run systematic fittings (e.g., $\chi^{2}$ fittings or MCMC) and thus these models should be understood in a qualitative sense, as probable rather than unique solutions. Nevertheless, our SED models still provide some general understanding of the target sources and some physical implications.

\begin{figure*}
    \centering
    \includegraphics[width=\linewidth]{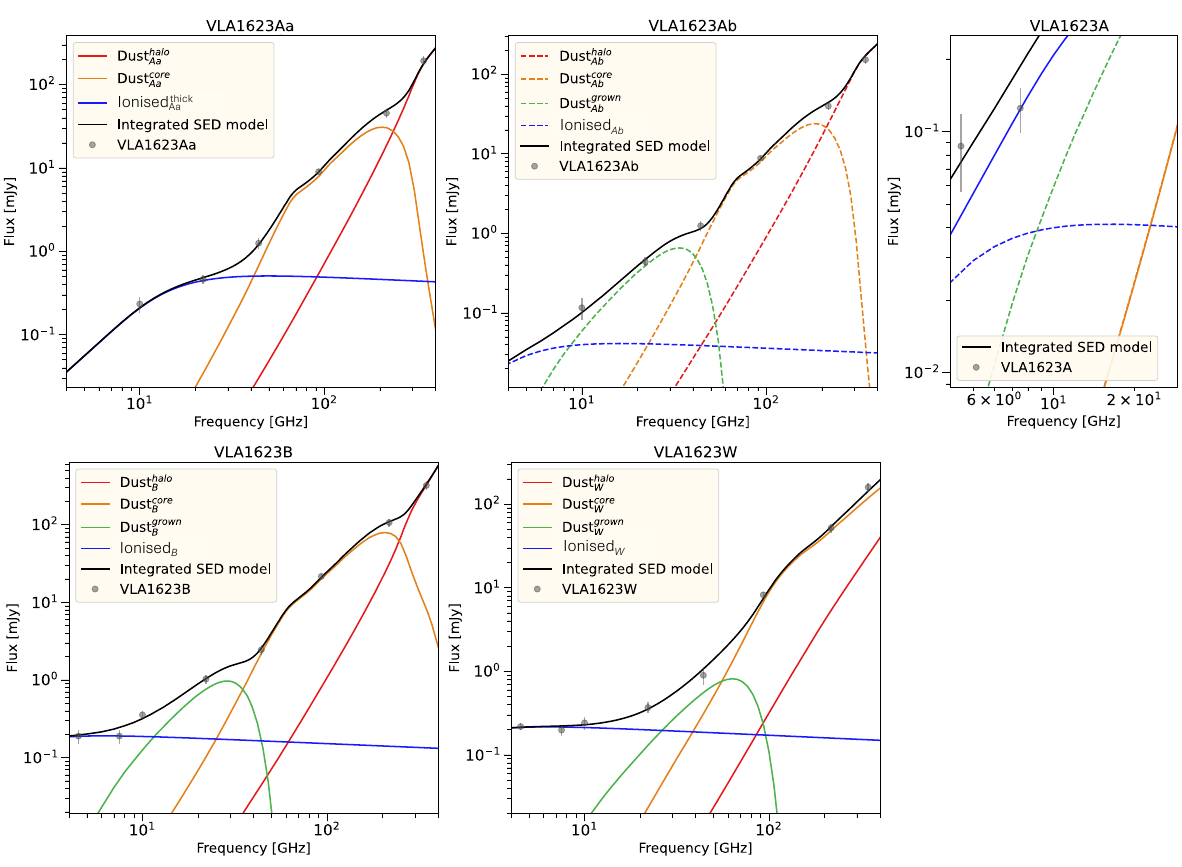}
    \caption{\textit{Top:} SED models for the VLA~1623~Aa and Ab resolved, integrated fluxes alongside the combined model, A, for the unresolved C-band data \citep{Dzib13}. Symbols show the observational data, coloured lines show the SEDs of individual emission components in our model (Table \ref{table:model}). For clarity we show the Aa emission components as solid lines and Ab components as dashed lines in the combined model. Integrated flux densities are shown in black lines. \textit{Bottom:} Same as the top but for resolved observations of VLA~1623~B and W respectively. 
    }
    \label{fig:1623SED}
\end{figure*}

We present our SED models for the four target sources in Figure~\ref{fig:1623SED} with the model parameters, summarised in Table \ref{table:model}. The names of the SED components we use in the model are for our convenience to reference them in text and do not have strict physical meanings. Below we discuss the specific fitting methodology used for each source.

\begin{deluxetable*}{ lccccccr }[ht!]
\tablecaption{Model parameters that reproduce the observed SEDs.\label{table:model}}
\tablehead{
\colhead{Component} &
\colhead{$T$} &
\colhead{$\Omega$\tablenotemark{a}} &
\colhead{$\Sigma_{\mbox{\tiny dust}}$} &  
\colhead{$a_{\mbox{\tiny max}}$} &
\colhead{EM} &
\colhead{Obscured by} &
\colhead{$M_{\mbox{\tiny dust}}$}
\\
\colhead{} &
\colhead{(K)} &
\colhead{($10^{-12}$ sr)} & 
\colhead{(g\,cm$^{-2}$)} &
\colhead{(mm)} &
\colhead{(cm$^{-6}$\,pc)} &
\colhead{} &
\colhead{$M_{\oplus}$}
\\
 & (1) & (2) & (3) & (4) & (5) & (6) & (7)
} 
\startdata
\multicolumn{8}{c}{VLA~1623~Aa} \\
Halo  & 170 & 0.63 & 0.25 & 0.2 & $\cdots$ & & 4.4 \\
Core  & 200 & 0.32 & 6.0 & 0.8 & $\cdots$ & Halo & 52 \\
Ionised gas  & 8000 & 9$\times$10$^{-3}$ & $\cdots$ & $\cdots$ & 8$\times10^{8}$ & & $\cdots$ \\
& & & & & & & Total: 56.4\\\hline
\multicolumn{8}{c}{VLA~1623~Ab} \\
Halo  & 170 & 0.5 & 0.4 & 0.2 & $\cdots$ & & 5.4 \\
Core  & 240 & 0.27 & 6.0 & 0.8 & $\cdots$ & Halo & 45 \\
Grown & 340 & 0.2 & 20 & 15 & $\cdots$ & Halo$+$Core & 110 \\
Ionised gas  & 8000 & 7$\times$10$^{-3}$ & $\cdots$ & $\cdots$ & 7.5$\times10^{7}$ & & $\cdots$ \\
& & & & & & & Total: 160\\\hline
\multicolumn{8}{c}{VLA~1623~B} \\
Halo  & 120 & 2.2 & 0.15 & 0.22 & $\cdots$ & & 9.1 \\
Core  & 153 & 1.18 & 6.0 & 1.0 & $\cdots$ & Halo & 194 \\
Grown & 180 & 0.8 & 20 & 15 & $\cdots$ & Halo$+$Core & 440 \\
Ionised gas  & 8000 & 0.22 & $\cdots$ & $\cdots$ & 1.0$\times10^{7}$ & & $\cdots$ \\
& & & & & & & Total: 643\\\hline
\multicolumn{8}{c}{VLA~1623~W} \\
Halo  & 20 & 3.5 & 0.2 & 0.05 & $\cdots$ & & 19 \\
Core  & 20 & 4.0 & 3.5 & 0.4 & $\cdots$ & Halo & 384 \\
Grown & 120 & 0.2 & 20 & 15 & $\cdots$ & Halo$+$Core & 110 \\
Ionised gas  & 8000 & 0.25 & $\cdots$ & $\cdots$ & 1.0$\times10^{7}$ & & $\cdots$ \\
& & & & & & & Total: 513\\\hline
\enddata
\tablecomments{
Similar to Table \ref{table:modelsingledust} but including multiple, mutually obscuring dust components for each source. In column (6) we include the components that are obscuring the emission of this component.
}
\end{deluxetable*}

\paragraph{\textbf{VLA~1623~A}}
\, The binary components, Aa and Ab, were spatially resolved between 10 and 217\,GHz (see Figure~\ref{fig:Gallery}). However, they were not resolved in the previous VLA C band (4.5 GHz and 7.5 GHz) observations \citep{Dzib13}.  The intra-band spectral index ($\alpha_{\text{7.5-4.5\,GHz}}$) is $\sim$0.7, which is consistent with ionised gas emission (see Appendix \ref{appendix:freefree}).  In at least one of the two binary components, ionised gas emission partly contributes to the integrated SEDs.  The fractional contribution of ionised gas emission may be higher in Aa, as Aa is considerably brighter than Ab in the resolved 10\,GHz continuum image.

At frequencies $>$\,20 GHz, the spectral shapes and flux densities of Aa and Ab are similar (Figure~\ref{fig:1623SED}). Their 350--217\,GHz spectral indices are $\sim$3, while the spectral indices between 22 GHz and 217 GHz are close to 2. In both binary components, the 350\,GHz flux densities are likely contributed by dust thermal emission from the relatively spatially extended, marginally optically thin halos (hereafter the `halo' components), which may represent the outer disks or inner envelopes. At $\sim$20--200 GHz, the flux densities may be partly or largely contributed by the thermal emission of the relatively spatially compact, optically thick disk cores (hereafter the `core' components).

The dust temperatures and solid angles of the halo and core components in Aa and Ab were (loosely) constrained thanks to the high angular resolutions of the 350 and 217\,GHz observations.  We approximated the solid angles of the halos components according to the deconvolved size scales at 350\,GHz \citep[see Table 2,][]{Harris18}, and adjusted the dust temperatures to make the flux densities consistent with the observations at 350 and 217\,GHz. The `core' components need to be obscured by the `halo' components in the line of sight to suppress the contribution of the `core' components at $>$217 GHz.  Otherwise, in the integrated SEDs, the strong contribution from the `core' components will yield 350-217\,GHz spectral index values that are too low to be consistent with observations (Figure~\ref{fig:1623SED}). Therefore, we make the $T_{\text{dust}}$/$\Omega_{\text{dust}}$ slightly larger/smaller than those of the `halo' components, while we note that there is a degeneracy between these two parameters. 

The $\Sigma_{\text{dust}}$ of the `halo' components were constrained by the 350--217\,GHz spectral indices. The $a_{\text{max}}$ in the `halo' components were not well constrained. We found that $a_{\text{max}}=200\,\mu$m may be a good solution, which provides modest extinction opacity at the required emission optical depths. 

The 217--93\,GHz spectral indices provided the lower limits for the $\Sigma_{\text{dust}}$ of the `core' components. The $a_{\text{max}}$ of the `core' components were once again barely constrained.  We found that choosing $a_{\text{max}}\sim0.8$\,mm helps maintain the constant $\sim$2 spectral indices between 44\,GHz and 217\,GHz, and provide sufficient extinction opacity at lower frequencies (explained below).  Fixing $a_{\text{max}}$ to 0.8\,mm, we adjusted the $\Sigma_{\text{dust}}$ of the `core' components to make them optically thick at $\gtrsim$50 GHz (Figure~\ref{fig:1623SED}).  In this case, the dust masses in the `core' components are $\sim$50 $M_{\oplus}$, which may be regarded as lower limits (Table \ref{table:model}). Operationally, we can continue increasing the $\Sigma_{\text{dust}}$ of the `core' components until their flux densities meet those observed at 10 and 22\,GHz. In that case, the dust masses in the `core' components will be further enhanced by 1--2 orders or magnitudes, which is not necessarily realistic. 

\paragraph{\textbf{VLA~1623~Aa}}

\,The simplest model to interpret the bright 10--22 GHz emission in VLA~1623~Aa (and part, if not all, of the 4.5 GHz and 7.5 GHz emission in the binary system) is with an ionised gas emission component that has a turnover frequency around 10--22 GHz (see the top left panel of Figure~\ref{fig:1623SED}; Table~\ref{table:model}; c.f. Appendix~\ref{appendix:freefree}). However, the parameters are uncertain as the C band measurements were taken well before our VLA observations, and we are not yet certain about the radio variability of this source (e.g., \citealt{Dzib13,Liu2014ApJ...780..155L,Coutens19}).  

In addition, we note that the lower limit of the 10--7.5\,GHz spectral index in Aa is $\sim$2.0.  Although this spectral index is consistent with optically thick ionised gas emission, there are currently no observations which can confirm whether or not an optically thick ionised gas emission source can exist in a low-mass young stellar object. The existing analytic models of ionized wind and jets which dominate the ionised gas emission in low-mass YSOs present surface density gradients.  Therefore, the spatially unresolved radio observations (e.g. 7.5\,--\,10\,GHz) will always simultaneously detect the optically thick wind/jet core with a spectral index $\sim$2.0, {in combination with the optically thin component that has a lower spectral index i.e. $<2$. In this case, the spectral indices of a relatively optically thick ionized wind/jet are expected to be $\sim$0.6, which is considerably lower than the lower limit constrained at 10--7.5\,GHz (\citealt{Wright1975MNRAS.170...41W,Reynolds86,Anglada1998AJ....116.2953A}). 

Due to the rationale discussed above, it may not be favourable to fully attribute the 10--22 GHz emission in source Aa to ionised gas emission. We discuss an alternative interpretation in Appendix \ref{appendix:AltModel} where part of the 10--22 GHz emission may be contributed by large embedded hot dust.

\paragraph{\textbf{VLA~1623~Ab}}

\,The 22-10\,GHz spectral index of source Ab is considerably higher than that in source Aa. Therefore, the 10 and 22\,GHz flux densities cannot be dominated by ionised gas emission.   We found that we can sufficiently reproduce the flux densities at the VLA bands by including a $T_{\text{dust}}=$\,340\,K component, namely the `grown' component (Table \ref{table:model}).  The $a_{\text{max}}$ is not constrained by our current observations.  We tentatively chose $a_{\text{max}}=$\,15\,mm to yield a high emissivity at 10 and 22\,GHz.  The dust mass of this component is $\sim$111 $M_{\oplus}$, which should regarded as a lower limit. 

\paragraph{\textbf{VLA 1623~B \& W}} 
\,Following the same methodology outlined above for VLA~1623~A, we find that the resulting SED models of VLA~1623~B and W are qualitatively similar to the model for VLA~1623~Ab. Both VLA~1623~B and W require a `grown' component to reconcile the fluxes at the VLA bands. However, in contrast to Aa, both B and W also require the inclusion of an optically thin ionised gas emission source to reproduce the observed spectral indices at low frequencies. This is due to their intra-band spectral indices ($\alpha_{\text{7.5-4.5\,GHz}}$) of $<-0.1$ \citep[see e.g.,][]{Dzib13}, which are inconsistent with dust emission alone.

\subsection{Embedding hot dust emission components}\label{appendix:AltModel}

Previous SED modelling analyses have shown that embedding high-temperature dust emission components that have relatively small $\Omega_{\text{dust}}$ can reproduce the bright emission and low spectral indices seen at low frequencies (\citealt{Li2017ApJ...840...72L,Liu2019ApJ...884...97L}).
By assuming millimetre or centimetre sized $a_{\text{max}}$ one can achieve bright emission at VLA bands. 

\begin{figure*}
    \centering
    \includegraphics[width=0.7\linewidth]{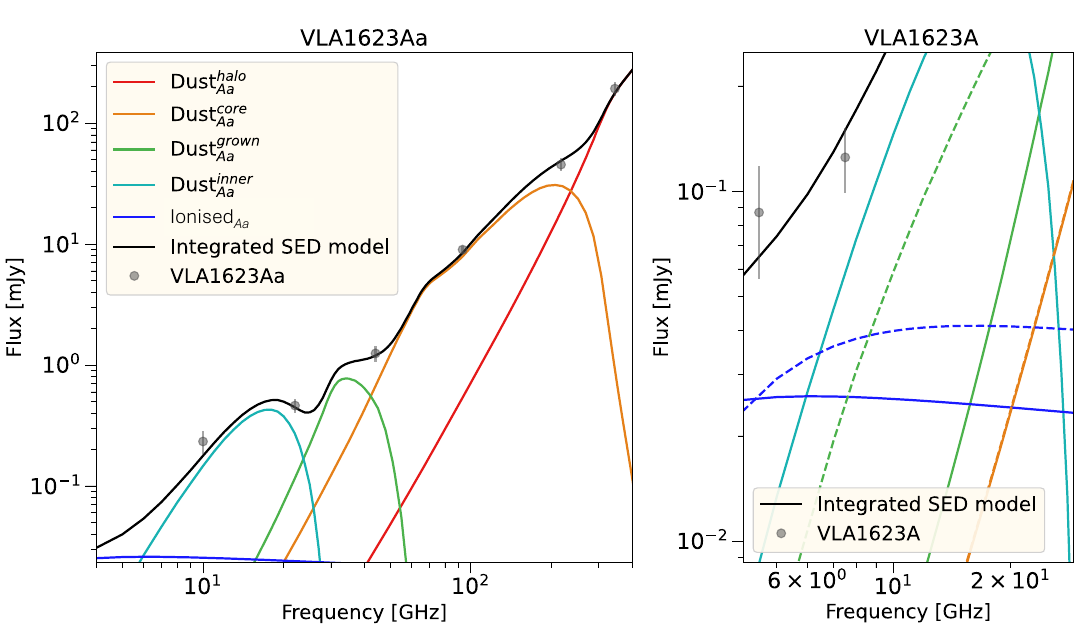}
    \caption{Embedded `hot dust' SED model for the VLA~1623~Aa resolved, integrated fluxes alongside the combined model for the unresolved C-band data \citep{Dzib13}. Symbols show the observational data, coloured lines show the SEDs of individual emission components in our model (Table \ref{table:modelAPX}). We show the SED model components for Ab as dashed lines in the combined model plot.}
    \label{fig:Aa_apx}
\end{figure*}

In Figure~\ref{fig:Aa_apx}, we present a model that can reproduce the observed flux densities by progressively embedding higher temperature and larger $a_{\text{max}}$ dust emission components on smaller scales (namely the `grown', and `inner' components, see Table \ref{table:modelAPX}).
Including the `grown' component and making it obscured by the `core' component allows the model to maintain the high flux density and flat spectral profile at 44--22\,GHz; including the `inner' component and making it obscured by the `grown' component allows maintaining the high flux density and flat spectral profile at 22--10\,GHz.
We note that the wiggles in the SED model are artificial and are a result of the discrete emission components we use to approximate the SED.
These wiggles can be eliminated by adopting an analytic model that has continuous density and temperature profiles, which is beyond the scope of the present paper. 

In the present model, the $a_{\text{max}}$ of the `core', `grown', and `inner' components were chosen such that the dust emission components have high absorption opacities and modest extinction opacities at 22\,GHz and 10\,GHz. 
In this case, the model provides the relatively conservative dust mass estimates for the embedded dust emission components, which may be regarded as lower limits (Table \ref{table:model}).

The $\Omega_{\text{dust}}$, $a_{\text{max}}$, $T_{\text{dust}}$, and $\Sigma_{\text{dust}}$ are degenerate due to the limited independent observational constraints. 
Nevertheless, we note that assuming smaller dust sizes (e.g., $a_{\text{max}}\lesssim$100 $\mu$m) will yield considerably higher dust masses, which are not necessarily realistic.
We also note that the upper limit of $T_{\text{dust}}$ is the $\sim$1,500 K sublimation temperature. We summarise the probable model parameters in table \ref{table:modelAPX}.

\begin{deluxetable*}{ lccccccr }[ht]
\tablecaption{Model parameters\label{table:modelAPX}}
\tabletypesize{\footnotesize}
\tablehead{
\colhead{Component} &
\colhead{$T$} &
\colhead{$\Omega$\tablenotemark{a}} &
\colhead{$\Sigma_{\mbox{\tiny dust}}$} &  
\colhead{$a_{\mbox{\tiny max}}$} &
\colhead{EM} &
\colhead{Obscured by} &
\colhead{$M_{\mbox{\tiny dust}}$}
\\
\colhead{} &
\colhead{(K)} &
\colhead{($10^{-12}$ sr)} & 
\colhead{(g\,cm$^{-2}$)} &
\colhead{(mm)} &
\colhead{(cm$^{-6}$\,pc)} &
\colhead{} &
\colhead{$M_{\oplus}$}
\\
 & (1) & (2) & (3) & (4) & (5) & (6) & (7)
} 
\startdata
\multicolumn{8}{c}{VLA~1623~Aa} \\
Halo  & 170 & 0.63 & 0.25 & 0.2 & $\cdots$ &  & 4.4 \\
Core  & 200 & 0.32 & 6.0 & 0.8 & $\cdots$ & Halo & 53 \\
Grown & 400 & 0.27 & 8.0 & 1.8 & $\cdots$ & Halo$+$Core & 60 \\
Inner & 1200 & 0.2 & 10 & 25 & $\cdots$ & Halo$+$Core$+$Grown & 55 \\
Ionised gas  & 8000 & 3$\cdot$10$^{-2}$ & $\cdots$ & $\cdots$ & 1$\cdot10^{7}$ & & $\cdots$ \\
& & & & & & & Overall: 172\\\hline
\enddata
\tablecomments{
(1) Dust temperature for dust emission components and electron temperature for the ionised gas emission component. (2) Solid angle of the emission components. (3) Dust column density. (4) Maximum (dust) grain size. (5) Emission measure for the ionised gas emission component. (6) The components that are obscuring the emission of this component. (7) Dust mass in units of Earth mass ($M_{\oplus}$).
}
\end{deluxetable*}

\end{document}